\documentclass[article,twocolumn,floats,nofootinbib,nobibnotes,superscriptaddress,10pt]{revtex4-1}

\usepackage{graphicx,amsmath,amssymb,hyperref}
\hypersetup{colorlinks=true, linkcolor=red, citecolor=magenta, urlcolor=blue}
\usepackage{dcolumn}
\usepackage{multirow}
\usepackage{xcolor,colortbl}
\usepackage{enumitem}
\usepackage{eqnarray}
\usepackage{makecell}
\usepackage{cancel}
\usepackage{etoolbox}
\usepackage{tabularx}
\usepackage{utfsym}

\renewcommand{\arraystretch}{1.5}

\begin{document}
\title{Constraining $z\lesssim 2$ ultraviolet emission \\ with the upcoming ULTRASAT satellite}

\author{Sarah Libanore}%
\email{libanore@bgu.ac.il}
\affiliation{Department of Physics, Ben-Gurion University of the Negev, Be'er Sheva 84105, Israel}

\author{Ely D. Kovetz}
\email{kovetz@bgu.ac.il}
\affiliation{Department of Physics, Ben-Gurion University of the Negev, Be'er Sheva 84105, Israel}

\begin{abstract}
The Extragalactic Background Light (EBL) carries a huge astrophysical and cosmological content: its frequency spectrum and redshift evolution are determined by the integrated emission of unresolved sources, these being galaxies, active galactic nuclei, or more exotic components. The near-UV region of the EBL spectrum is currently not well constrained, yet a significant improvement can be expected thanks to the soon-to-be launched Ultraviolet Transient Astronomy Satellite (ULTRASAT). Intended to study transient events in the 2300-2900\,\AA\,observed band, this detector will provide 
wide field maps, tracing the UV intensity fluctuations on the largest scales. 
In this paper, we suggest how to exploit ULTRASAT full-sky map, as well as its low-cadence survey,
to reconstruct the redshift evolution of the UV-EBL volume emissivity. We build upon the work of Chiang et al.\ (2019), where the Clustering-Based Redshift (CBR) technique was used to study diffuse light maps from GALEX. Their results showed the capability of the cross correlation between GALEX and SDSS spectroscopic catalogs in constraining the UV emissivity, highlighting how CBR is sensitive only to the extragalactic emissions,  avoiding foregrounds and Galactic contributions. 
In our analysis, we introduce a framework to forecast the CBR constraining power when applied to ULTRASAT and GALEX in cross correlation with the 5-year DESI spectroscopic survey.   
We show that these will yield a strong improvement in the measurement of the UV-EBL volume emissivity. For $\lambda = 1500\,$\AA\,non-ionizing continuum below $z \sim 2$, we forecast a $1\sigma$ uncertainty $\lesssim 26\%\,(9\%)$ with conservative (optimistic) bias priors using ULTRASAT full-sky map; similar constraints can be obtained from its low-cadence survey, which will provide a smaller but deeper map. We finally discuss how these results will foster our understanding of UV-EBL models.
\end{abstract}

\maketitle

\section{Introduction}

Across the whole electromagnetic spectrum, light from unresolved sources of astrophysical and cosmological origins overlaps, producing the so-called Extragalactic Background Light (EBL). Such integrated emission is observed and constrained through different probes; reviews in~Refs.~\cite{Cooray:2016jrk,Hill:2018trh} describe the main features of the EBL and highlight the degree of accuracy reached in the various frequency bands. Some regimes have been widely analyzed, for instance the microwave band where the cosmic microwave background (CMB) is observed or the optical-infrared region that collects redshifted emission from stars. 
Other frequency bands, instead, are still largely unprobed: the near-ultraviolet (NUV) to far-ultraviolet (FUV) region is one of these. Between $\nu_{\rm obs} \sim 1$\,-\,$3\cdot 10^6\,{\rm GHz}$ ($\lambda_{\rm obs}\sim 3000$\,-\,$1000\,$\AA) a lower EBL limit has been estimated both using the integrated light from galaxy number counts and through photometric measurements from the Voyager 1 and 2 missions~\cite{Murthy:1999,Eldstein:2000}, Hubble~\cite{Brown:2000uq,Bernstein:2001sq} and the Galaxy Evolution Explorer (GALEX,~\cite{Martin:2004yr,Morrissey:2007hv,Murthy:2010vf}), even if with large statistical and systematic errors~\cite{Cooray:2016jrk}.
This lack of measurements is not only due to technological issues: extragalactic photons in this regime get mostly absorbed by neutral hydrogen in the intergalactic medium (IGM) and inside the Milky Way. Moreover, Galactic light in this part of the spectrum provides a strong foreground contribution. This problem also affects other wavebands, for example the optical range; techniques have been developed to characterize and remove the foregrounds (see, e.g.,~the measurements from the New Horizon mission in~Ref.~\cite{Symons:2022lke}).

The study of the UV background plays an important role in understanding structure formation and star formation history (depicted e.g.,~in~Refs.~\cite{Madau:2014bja,Bernstein:2001sq} and references within). First of all, its spectral energy distribution provides an unbiased estimate of the energy produced by star formation. Moreover, the EBL can be used to study the chemical enrichment history of the Universe, the total baryon fraction in stars and the local mass density of metals. 
The background radiation in this frequency band is highly correlated with the one in the infrared range~\cite{Sasseen:1995,Schiminovich:1999qb}, which is consistent with an origin related with dust-scattered stellar radiation. The UV-EBL can also contain contributions of cosmological origin, such as black hole accretion in quasars and active galactic nuclei (AGN), direct-collapse black hole formation or decaying and annihilating dark matter (see e.g.,~Refs.~\cite{Dwek:1998bk,Bond:1985pc,Yue:2012dd,Henry:2014jga,Creque-Sarbinowski:2018ebl,Kalashev:2018bra,Bernal:2020lkd,Carenza:2023qxh}). 

Improving our knowledge of the UV-EBL is therefore a goal that should be pursued with upcoming instruments and surveys. In this context, a very interesting opportunity will soon be offered by the Ultraviolet Transient Astronomy Satellite (ULTRASAT,~\cite{Sagiv:2013rma,ULTRASAT:2022,Shvartzvald:2023ofi}), whose launch is scheduled for 2026.
ULTRASAT foresees as its primary goal the study of astrophysical transients in the NUV band. In order to achieve this, part of the first six months of the mission will be dedicated to the realization of a full sky map,\footnote{The map will use an integration time of $15,000$ seconds at Galactic latitudes $|b| > 30\,$deg. See the~\href{https://www.weizmann.ac.il/ultrasat/}{ULTRASAT official website} for more details.} to be used as a reference frame for subsequent observations. Such map will reach a limiting AB magnitude of 23.5. 
Moreover, during its regular science operation, ULTRASAT will scan repeatedly $\sim 40$ selected extragalactic fields, in the so called {\it low-cadence survey} (LCS,~\cite{Shvartzvald:2023ofi}).\footnote{In the low-cadence survey, ULTRASAT plans to observe 40 extragalactic fields with 900s integration time each; over the full duration of the mission, this will result in a theoretical limiting magnitude $\sim 24.5$\,-\,$25.5$. }
Preliminary studies on the PSF and on the expected confusion noise, suggest that these images will provide a $\sim 6800\,{\rm deg}^2$ map with magnitude limit $\sim 24.5$. 

These outputs will provide a $\sim 10$\,-\,$100$ times sensitivity improvement with respect to the currently existing benchmark, namely the diffuse UV light map constructed in ~Ref.~\cite{Murthy:2014}.

This work used data from the All Sky and Medium Imaging Surveys performed by GALEX, masking resolved sources and binning pixels in order to estimate the combined contribution of the UV-EBL~\cite{Sujatha:2009,Sujatha:2010fd} and foregrounds, these being sourced by Galactic dust-scatter light, near-Earth airglow, or other~\cite{Hamden:2013,Murthy:2013,Henry:2014jga,Akshaya:2018}; the origin of part of the foreground is yet unidentified.
The GALEX map provided important results on the study of dust scattering emissions (e.g.,~Refs.~\cite{Murthy:2016,Chiang:2019}), correlation between UV and IR emissions (e.g.,~Refs.~\cite{Murthy:2014,Saikia:2017}), and extragalactic sources (e.g.,~Refs.~\cite{Welch:2020,Chiang:2018miw}). The frequencies probed by GALEX allowed to access the UV-EBL between redshifts $z\!\sim\! 0-1.5$; ULTRASAT will allow us to improve the sensitivity around $z\sim 1$, pushing the observations towards slightly higher $z$. 

An innovative application has been presented by the authors of~\hypertarget{C19}{\cite{Chiang:2018miw}} (\hyperlink{C19}{\color{magenta} C19} throughout this paper). UV satellites as GALEX and ULTRASAT observe a wide frequency range; therefore, radiation with various rest frame wavelengths, emitted at different cosmological distances from the observer, can be redshifted inside their observational band. 
The redshift dependence of the integrated, observed emission can be reconstructed through {\it broadband tomography}, which is based on the cross correlation between UV data and spectroscopic galaxy samples. This technique, also referred to as \textit{Clustering-Based Redshift} (CBR), was originally developed in~Refs.~\cite{Newman:2008mb,McQuinn:2013ib,Menard:2013aaa}, and applied in different contexts, for instance to obtain a statistic estimate of the redshift of photometric samples~\cite{Rahman:2014lfa,Scottez:2016fju,Morrison:2016stl,DES:2017rfw,vandenBusch:2020lur}, to study the cosmic infrared background~\cite{Schmidt:2014jja,Cheng:2021wex}, or the presence of extragalactic sources in Galactic dust maps~\cite{Chiang:2019}, and to improve the constraining power of radio surveys on cosmological parameters~\cite{Kovetz:2016hgp,Scelfo:2021fqe}. Cross correlation between EBL and galaxy surveys have also been used to test star formation models~\cite{Sun:2023ewx} and to the derive luminosity and mass functions for photometric galaxy catalogs~\cite{Bates:2018lxv}. \hyperlink{C19}{\color{magenta} C19} applied this technique to the GALEX map, using catalogs from the Sloan Digital Sky Survey (SDSS,~\cite{SDSS:2004dnq,
Reid:2015gra}), to reconstruct the redshift evolution of the UV-EBL volume emissivity. A follow-up of this work was realized in~\hypertarget{S21}{\cite{Scott:2021zue}} (\hyperlink{S21}{\color{magenta} S21} in the paper), where the authors applied a similar formalism to forecast the constraining power of the  Cosmological Advanced Survey Telescope for Optical and UV Research (CASTOR,~\cite{Cote:2019}) in cross correlation with Spectro-Photometer for the History of the Universe, Epoch of Reionization and Ices Explorer (SPHEREx,~\cite{SPHEREx:2014bgr,SPHEREx:2016vbo,SPHEREx:2018xfm}). 

The goal of our work is to build on the \hyperlink{C19}{\color{magenta} C19} analysis, applying it to the forecasted capabilities of ULTRASAT in cross correlation with upcoming spectroscopic galaxy surveys, in particular the Dark Energy Spectroscopic Instrument (DESI,~\cite{DESI:2013agm,DESI:2016fyo,DESI:2016igz}). When the full release of DESI data will be available, it will also be possible to cross correlate its spectroscopic catalogs with the GALEX map from~Ref.~\cite{Murthy:2014}, and further improve the results from  \hyperlink{C19}{\color{magenta} C19}. In our analysis, therefore, we consider the possibility of combining together GALEX and ULTRASAT maps with DESI, to perform a powerful CBR analysis, to access the UV-EBL particularly around $z\sim 1$. 

Our paper is structured as follows. We begin in Sect.~\ref{sec:ultrasat} by presenting the ULTRASAT satellite and giving some information about its specifications, in particular with respect to the full-sky map and the estimate of their 
noises. The approach we follow is inspired by line-intensity mapping science (see e.g.,~Refs.~\cite{Kovetz:2017agg,Bernal:2022jap} for review). We proceed in Sect.~\ref{sec:modeling} by modeling the signal that we aim to constrain, namely the UV volume emissivity (\ref{sec:emissivity}). We then build the observable of interest in the context of the CBR technique, namely the angular cross correlation between an intensity map and a spectroscopic galaxy survey (\ref{sec:intensity} and~\ref{sec:CBR}), for which we also provide a noise estimate (\ref{sec:noise}). Section~\ref{sec:analysis} presents our forecast analysis and collects our results. In particular, we start by reproducing the \hyperlink{C19}{\color{magenta} C19} constraints for GALEX$\times$SDSS (\ref{sec:galex_post}), and then we proceed by applying the technique to ULTRASAT$\times$DESI and (GALEX$+$ULTRASAT)$\times$DESI, comparing our forecasted results for the volume UV-EBL emissivity (\ref{sec:results_emissivity}) with various constraints available in the literature. We show that this approach will yield a strong improvement in the measurement of the UV-EBL volume emissivity. Our conclusions are presented and discussed in Sect.~\ref{sec:Conclusion}.

\section{The ULTRASAT satellite}\label{sec:ultrasat}

The observational window of ULTRASAT will span between $\lambda_{\rm min} = 2300\,$\AA, $\lambda_{\rm max} = 2900\,$\AA, which implies an observed frequency range centered at $\nu_{\rm obs} = 1.17\cdot 10^6\,{\rm GHz}$.
To help understanding which redshifts can be mapped using these wavelengths, we consider Lyman-$\alpha$ (Ly$\alpha$) line emission with rest frame $\lambda_{\rm Ly\alpha} = 1216\,$\AA: provided that $\lambda_{\rm obs} = \lambda_{\rm rest}(1+z)$, ULTRASAT will observe $z_{\rm Ly\alpha} \in [0.9,1.4]$. 

In our analysis in Sect.~\ref{sec:analysis}, we assume to rely on the UV map obtained during the first six months of the ULTRASAT mission. We assume that these will be realized using a technique similar to the one~Ref.~\cite{Murthy:2014} applied to GALEX data. The dataset we assume to analyze, hence consists of UV intensity values measured in the pixels of each map; the map is characterized by an intensity average value, which we also refer to as the {\it monopole}, and by spatial fluctuations; this indeed makes its study analogous to what is done in line-intensity mapping surveys~\cite{Kovetz:2017agg,Bernal:2022jap}. 

For both ULTRASAT {full-sky map} and GALEX, the observed field is almost full-sky. The pixel scale is $5.4''/{\rm pix}$ for ULTRASAT, $5''/{\rm pix}$ for GALEX. 
The response function $R(\lambda_{\rm obs})$ of the instruments is defined in terms of the quantum efficiency, that is the number of electrons per incident photon depending on the frequency, normalized through
\begin{equation}\label{eq:response}
    \int_{\lambda_{\rm obs}^{\rm min}}^{\lambda_{\rm obs}^{\rm max}}d\lambda_{\rm obs}\frac{R(\lambda_{\rm obs})}{\lambda_{\rm obs}} = 1.
\end{equation}
We compute the ULTRASAT response function based on information in~Refs.~\cite{Bastian_Querner_2021,Asif:2021vrm}, and we show it in Fig.~\ref{fig:model_summary}. Here, we also show $R(\lambda_{\rm obs})$ for the two GALEX filters, respectively centered in the NUV range $\lambda_{\rm obs} \in [1750,\,2800]\,$\AA\,and in the FUV range $\lambda_{\rm obs} \in [1350,\,1750]\,$\AA.\footnote{GALEX response functions are provided in the public repository \url{http://svo2.cab.inta-csic.es/svo/theory/fps3/}.}  GALEX NUV filter observes Ly$\alpha$ in a similar range with respect to ULTRASAT, while GALEX FUV refers to a lower redshift regime,~$z_{\rm Ly\alpha} \in [0.1,0.4]$. GALEX FUV and NUV filters, therefore, are complementary in probing Ly$\alpha$ emissions in the local Universe; ULTRASAT will improve the sensitivity mostly at $z\sim 1$. 

The signal of interest for our analysis is modeled in the next section; here instead we provide an estimate of the noise variance in the maps. 
As \hyperlink{C19}{\color{magenta} C19} details, masking the pixels with flux above the detection limit, resolved sources can be separated from the diffuse light contribution. This separation on the map level helps to implement the algorithms required to mitigate the foreground noise; on the other hand, a combined analysis of the two components allows us to study the UV intensity field without selection effects. From now on, we thus assume that a procedure similar to \hyperlink{C19}{\color{magenta} C19} has been performed: resolved sources and diffuse light are initially separated to create the maps and estimate their noises, and then summed to perform the analysis.

The noise variance per pixel in the ``overall" map is estimated based on the $5\sigma$ AB magnitude limit $m_{\rm AB}$, weighted by the different exposure times during the observations. In the case of GALEX NUV and FUV, \hyperlink{C19}{\color{magenta} C19} indicates $m_{\rm AB}\sim 20.5$\,-\,$23.5$; we adopt the intermediate value $m_{\rm AB}=22$.\footnote{The value is chosen consistently with analysis in \hyperlink{C19}{\color{magenta} C19}. We thank Y.~K.~Chiang for discussion.} For ULTRASAT reference map  instead, following~Ref.~\cite{Shvartzvald:2023ofi}, we assume $m_{\rm AB }=23.5$, so to get a $\times 10$ sensitivity improvement from GALEX. 

The limiting magnitude characterizes the detection level above which sources (with Galactic or extragalactic origin) cannot be resolved; for this reason, we can use it to describe the ground level of the observed intensity inside a region with comparable size to the point spread function. We convert $m_{\rm AB}$ to flux per pixel via $m_{\rm AB} = 8.90 - 2.5\log_{10}\mathcal{F}_{\nu}$, and we rescale it to flux density inside a certain angular region as $f_{\nu}^{\rm pix} = \mathcal{F}_{\nu}/A_{\rm pix}$. The pixel surface area is computed as $A_{\rm pix} = L_{\rm pix}^2$, where $L_{\rm pix}$ is the pixel scale. 
The value obtained corresponds to sources 
$5$ times brighter than the noise level; hence, we get the total noise variance per pixel 
\begin{equation}\label{eq:sigma_N}
    \sigma_N^2 = \left[{f_\nu^{\rm pix}}/{5}\right]^2 .
\end{equation}
To reduce the value of $\sigma_N^2$, it is always possible to combine the pixels in groups, which we call ``effective pixels": this reduces their number, and it averages noise fluctuation on the smallest scales. We will discuss in Sect.~\ref{sec:noise} that, in the context of the CBR analysis, \hyperlink{C19}{\color{magenta} C19} used effective pixels with $L_{\rm pix}^{\rm eff} = 50\,''/{\rm pix}$; we will show that instead, ULTRASAT will obtain good results already using $L_{\rm pix}$. Note that, so far, we did not account for foreground contributions; these will be described in Sect.~\ref{sec:noise}.

Table~\ref{tab:surveys} collects the main survey specifications we discussed so far, and it compares their estimated $\sigma_N^2$ value. The table also reports the value of the observed monopole $J_{\nu_{\rm obs}}$, which we define in Sect.~\ref{sec:intensity}.
\begin{table}[ht!]
    \centering\small
    \addtolength{\tabcolsep}{-0.35em}
    \begin{tabular}{|c|ccc|cc|}
    \hline
     & $\lambda_{\rm obs}$ &  \multirow{2}{*}{$z_{\rm Ly\alpha}$}  & $L_{\rm pix}^{\rm (eff)}$  & $J_{\nu_{\rm obs}}$& $\sigma_N^2$ \\
     & [{\AA}] & & [``/px] & [{\rm Jy/sr}] & [${\rm Jy^2/sr^2}$] \\
    \hline
    ULTRASAT  & {[2300,\,2900]} & {[0.9, 1.4]} & {5.45} & {264} & 422 \\
    \hline
     \multirow{2}{*}{GALEX}  & [1350,\,1750] & [0.1, 0.4] & \multirow{2}{*}{5~(50)} & 237 & \multirow{2}{*}{1959} \\
      & [1750,\,,2800]   & [0.4, 1.3] &  & 79 &   \\
    \hline
    \end{tabular}
    \smallskip 
    \caption{Specs of the broadband surveys ULTRASAT~\cite{Shvartzvald:2023ofi} and GALEX~\cite{Martin:2004yr}. The redshift range $z_{\rm Ly\alpha}$ is computed for the Ly$\alpha$ line $\lambda_{\rm Ly\alpha} = 1216\,$\AA; other parts of the UV spectrum are shifted to different ranges. We indicate both the pixel scale $L_{\rm pix}$ and (in parenthesis) the effective scale of the pixels used in the analysis in Sect.~\ref{sec:analysis}. The monopole $J_{\nu_{\rm obs}}$ and the variance $\sigma_N^2$ are estimated using respectively Eqs.~\eqref{eq:intensity} and~\eqref{eq:sigma_N}.}
    \label{tab:surveys}
\end{table}

\section{Modeling the signal and noise}\label{sec:modeling}

The broadband measurements ULTRASAT will realize integrate the radiation emitted over a wide frequency and redshift range. Reconstructing the spectral shape of the signal, as well as its evolution across cosmic time, is of fundamental importance to disentangle cosmological dependencies and astrophysical properties of the emitters. One way this goal can be pursued is via the broadband tomography, or clustering-based redshift (CBR) technique~\cite{Newman:2008mb,McQuinn:2013ib,Menard:2013aaa}, namely the cross correlation between one broadband survey (ULTRASAT or GALEX, in our case) and a reference spectroscopic galaxy catalog. Using cross correlations allows us to overcome 
the presence of Galactic foregrounds, ensuring that the measured signal has extragalactic origin. The possibility of chance correlation is further reduced by the fact that foregrounds fluctuate mainly at large scales, while we are interested in cross correlations on small scales. Hence, in our analysis, foregrounds only contribute to the noise budget.

The authors of \hyperlink{C19}{\color{magenta} C19} applied CBR to GALEX data, in order to constrain the redshift evolution of the UV-EBL volume emissivity. In their case, a subset of the SDSS catalogs~\cite{SDSS:2004dnq,
Reid:2015gra} was adopted as reference. 
ULTRASAT, with its improved NUV capability, seems to be the perfect heir for this kind of analysis. Considering its timeline, a bunch of state-of-the-art galaxy surveys will be available to perform the cross correlation. As we will detail in Sect.~\ref{sec:noise}, the best results are obtained for small spectroscopic redshift uncertainties and large galaxy number densities in the range under analysis. For this reason, we decided to rely on the DESI~\cite{DESI:2013agm,DESI:2016fyo,DESI:2016igz} forecasted 5\,years capability, which are depicted in~Ref.~\cite{DESI:2023dwi}. 

To forecast the constraining power ULTRASAT$\times$DESI will have compared to GALEX$\times$SDSS, we introduce the quantities required to model the observables.  

\noindent
In Sect.~\ref{sec:emissivity}, we follow the formalism from \hyperlink{C19}{\color{magenta} C19} and \hyperlink{S21}{\color{magenta} S21} and we describe the UV-EBL volume emissivity that we aim to constrain. The emissivity is then converted into observed intensity in Sect.~\ref{sec:intensity}. We then review how CBR can reconstruct its redshift evolution in Sect.~\ref{sec:CBR}, and we estimate the uncertainty in its measurement in Sect.~\ref{sec:noise}. Figure~\ref{fig:model_summary} summarizes the logic and modeling of the analysis.

\subsection{UV volume emissivity}\label{sec:emissivity}

The first ingredient we need in order to model the signal is the spatially averaged comoving volume UV emissivity, $\epsilon(\nu,z)$ $[\rm erg\,s^{-1}Hz^{-1}Mpc^{-3}]$. In analogy to \hyperlink{C19}{\color{magenta} C19} and \hyperlink{S21}{\color{magenta} S21}, we parameterize its frequency and redshift dependencies using a piecewise function, capable of describing the different spectral contributions the EBL receives. This model is in agreement with simulation results based on the study of the radiative transfer of Lyman continuum photons in the IGM~\cite{Haardt:2011xv}. 

We distinguish between the non-ionizing continuum at $\lambda_{\rm rest} > 1216\,$\AA\, and between $912\,$\AA $< \lambda_{\rm rest} \leq 1216\,$\AA, 
the Ly$\alpha$ line at $\lambda_{\rm rest} = 1216\,$\AA, and
the ionizing continuum at $\lambda_{\rm rest} \leq 912\,$\AA, below the Lyman break.
Above Ly$\alpha$, we describe the frequency and redshift evolution of the non-ionizing continuum with respect to a pivot value as 
\begin{equation}\label{eq:epsilon_1}
    \epsilon(\nu,z) =
        \epsilon_{1500}\left[\frac{\nu}{\nu_{1500}}\right]^{\alpha_{1500}},
\end{equation}
where $\nu_{1500}$ is the frequency at $\lambda_{\rm rest} = 1500\,$\AA. 

\noindent
Between the Ly$\alpha$ line and the Lyman break, we adopt 
\begin{equation}\label{eq:epsilon_2}
    \epsilon(\nu,z) = \epsilon_{1500}\left[\dfrac{\nu_{1216}}{\nu_{1500}}\right]^{\alpha_{1500}} \begin{cases}
        &\left[\dfrac{\nu}{\nu_{1216}}\right]^{\alpha_{1100}}\,\, \\
        & \dfrac{{\rm EW}}{\nu_{1216}\delta}\dfrac{\nu^2}{c} + \left[\dfrac{\nu}{\nu_{1216}}\right]^{\alpha_{1100}}
\end{cases}
\end{equation}
where the first line holds for $|\nu - \nu_{1216}| >\nu_{1216}\delta$, with $\delta = 0.005$ and line frequency $\nu_{1216} =  2.5 \times 10^{6}\,{\rm GHz}$. 
Finally, in the ionizing range we set 
\begin{equation}\label{eq:epsilon_3}
    \epsilon(\nu,z) =
        \epsilon_{1500}\left[\frac{\nu_{1216}}{\nu_{1500}}\right]^{\alpha_{1500}}f_{\rm LyC}\left[\frac{\nu_{912}}{\nu_{1216}}\right]^{\alpha_{1100}}\left[\frac{\nu}{\nu_{912}}\right]^{\alpha_{900}},
\end{equation}
where $f_{\rm LyC}$ represents the ionizing photons escape fraction, and it determines the presence of the Lyman break and Lyman forest in the spectra of galaxies and quasars~\cite{Meiksin:2003ue,Khaire:2018fqp}.

\begin{figure}[ht!]
    \centering     \includegraphics[width=\columnwidth,height=.8\textheight]{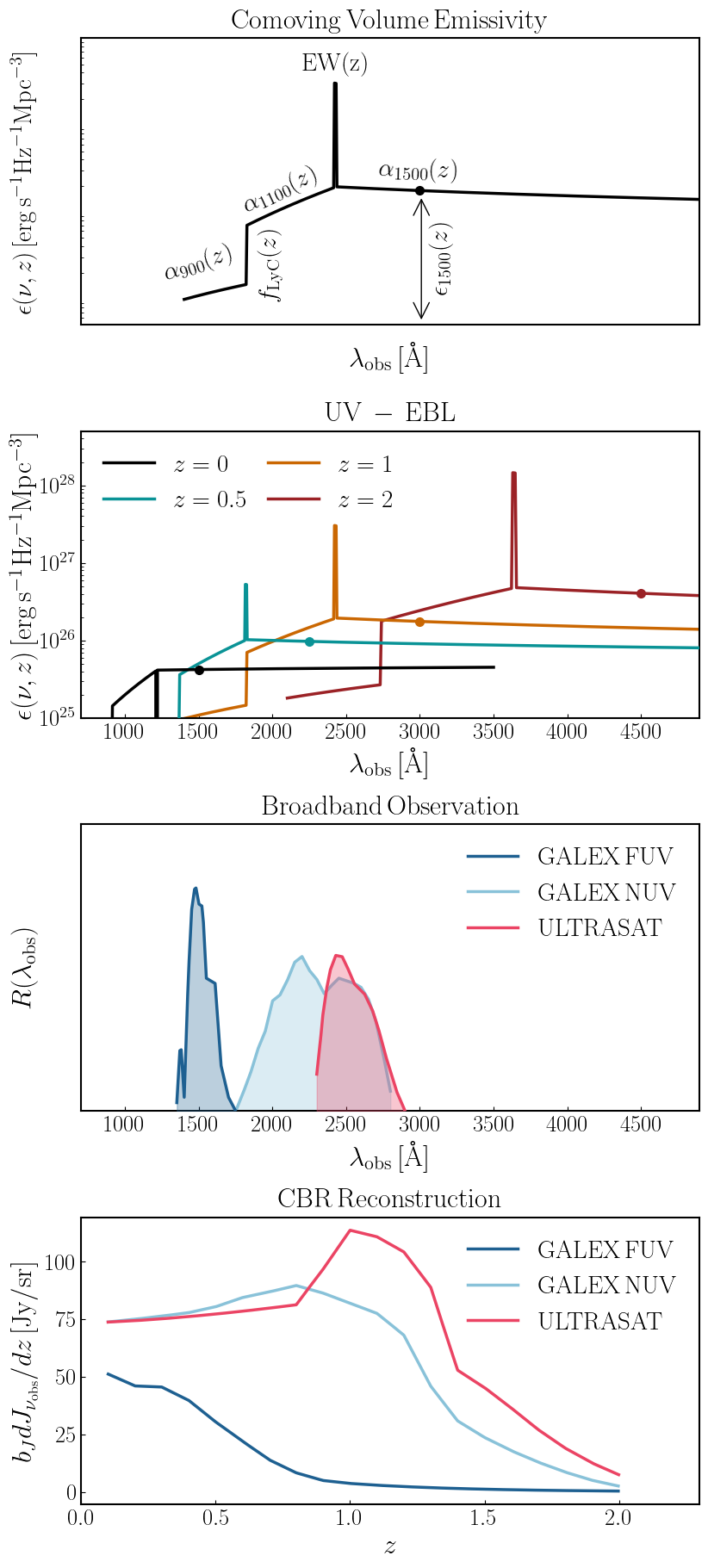}
    \caption{Summary of the modeling in our analysis. We characterize the UV volume emissivity $\epsilon(\nu,z)$ based on Eqs.~\eqref{eq:epsilon_1} to~\eqref{eq:parameters_zdep} (panel 1). Emissions from different $z$ contribute to the EBL (panel 2, the dots show $\lambda_{\rm rest}=1500\,$\AA; the Ly$\alpha$ line is in absorption at low $z$), which is weighted by the detector response function in the broadband observation in Eq.~\eqref{eq:intensity} (panel 3).
    The CBR allows us to reconstruct $b_J(z)dJ_{\nu_{\rm obs}}/dz$ in Eqs.~\eqref{eq:dJnu},~\eqref{eq:bias_J} (panel 4). Figure inspired by~Ref.~\cite{Chiang:2018miw}. }
    \label{fig:model_summary}
\end{figure}

The redshift dependencies of Eqs.~\eqref{eq:epsilon_1},~\eqref{eq:epsilon_2},~\eqref{eq:epsilon_3} are encapsulated in the normalization and slope parameters of the continuum, namely 
\begin{equation}\label{eq:parameters_zdep}
\begin{aligned}
    & \epsilon_{1500} = \epsilon_{1500}^{z=0}(1+z)^{\gamma_{1500}},\\
    & \alpha_{1500} = \alpha_{1500}^{z=0}+C_{1500}\log_{10}(1+z),\\
    & \alpha_{1100} = \alpha_{1100}^{z=0}+C_{1100}\log_{10}(1+z),\\
    & \alpha_{900} = \alpha_{900}^{z=0},\\
\end{aligned}
\end{equation}
as well as in the line equivalent width,\footnote{The equivalent width is usually defined as the side of a rectangle whose height is equal to the continuum emission and the area is the same as the line itself, namely ${\rm EW} = \int d\lambda (1-J_s(\lambda)/J_c(\lambda))$, where $J_c(\lambda)$ is the continuum intensity, while $J_s(\lambda) = J_l(\lambda)+J_c(\lambda)$ is the spectrum intensity, including both the continuum and the line. In our convention, a positive value of EW gives rise to an emission line, while a negative value produces an absorption line.}  
\begin{equation}\label{eq:EW}
\begin{aligned}
   & {\rm EW} = C_{\rm Ly\alpha}\log_{10}\left(\frac{1+z}{1+z_{\rm EW 1}}\right) + {\rm EW}^{z = z_{\rm EW 1}},\\
   &  C_{\rm Ly\alpha} = \frac{{\rm EW}^{z = z_{\rm EW 2}}-{\rm EW}^{z = z_{\rm EW 1}}}{\log_{10}[(1+z_{\rm EW 2})/(1+z_{\rm EW 1})]}\,,
\end{aligned}
\end{equation}
and in the Lyman continuum escape fraction in the IGM 
\begin{equation}\label{eq:fLyC}
\begin{aligned}
   & \log_{10}f_{\rm LyC} = C_{\rm LyC}\log_{10}\left(\frac{1+z}{1+z_{\rm C 1}}\right) + \log_{10}f_{\rm LyC}^{z = z_{\rm C 1}}\,,\\
   &  C_{\rm LyC} = \frac{\log_{10}f_{\rm LyC}^{z = z_{\rm C 2}}-\log_{10}f_{\rm LyC}^{z = z_{\rm C 1}}}{\log_{10}[(1+z_{\rm C 2})/(1+z_{\rm C 1})]}\,.
\end{aligned}
\end{equation}
Overall, as Fig.~\ref{fig:model_summary} shows in the top panel, the amplitude of the volume emissivity is set by $\epsilon_{1500}(z)$, while the characteristics of the Ly$\alpha$ line are determined by ${\rm EW}(z)$, and the amplitude of the Lyman break by $f_{\rm LyC}(z)$. At each $z$, the $\alpha_{l}(z)$ parameters $(l=1500,1100,900)$ set the slopes of the frequency dependence in the different regimes.

The upper part of Table~\ref{tab:pars} collects the parameters through which we parameterize the volume emissivity $\epsilon(\nu,z)$, together with their fiducial values, which we choose to match the posterior results of \hyperlink{C19}{\color{magenta} C19}. As for the $f_{\rm LyC}^{z_{\rm C1}},\,f_{\rm LyC}^{z_{\rm C2}}$ parameters, the analysis in \hyperlink{C19}{\color{magenta} C19} only provided 3$\sigma$ upper bounds; in the following, we hence set these parameters to the largest values allowed by their results. This is done to show the capability ULTRASAT will have in improving the constraints on such upper limits; in Sect.~\ref{sec:forecast_sources} we discuss the implications in the case they are set to smaller values.

To analyze both the GALEX NUV and FUV filters, \hyperlink{C19}{\color{magenta} C19} sets the pivot points of the EW in Eq.~\eqref{eq:EW} to $\{z_{\rm EW 1} = 0.3,\,z_{\rm EW 2}=1\}$, and the ones of the escape fraction in Eq.~\eqref{eq:fLyC} to $\{z_{\rm C 1}=1,\,z_{\rm C 2}=2\}$. In our ULTRASAT analysis, we decided to keep the same pivot points for the escape fraction, while we pivoted the EW at $\{z_{\rm EW 1} = 1,\,z_{\rm EW 2}=2\}$: the higher redshift has been chosen according to the ULTRASAT $\lambda_{\rm obs}$ regime, which collects emissions from the Ly$\alpha$ line at $z_{\rm Ly\alpha} \in [0.9,\,1.4]$ (compare with Table~\ref{tab:surveys}) and from the Lyman break at $z_{912} \in [1.5,\,2.2]$. To get a continuous signal, we estimated 
${\rm EW}^{z = 2}$ from Eq.~\eqref{eq:EW}, using the \hyperlink{C19}{\color{magenta} C19} pivots points.
Figure~\ref{fig:model_summary} shows $\epsilon(\nu,z)$ at different $z$. 
\begin{table}[ht!]
    \centering
\renewcommand{\arraystretch}{1.2}
\addtolength{\tabcolsep}{-0.2em}
    \begin{tabular}{|cc||cc|}
    \hline
     Parameter & Fiducial value &  Parameter & Fiducial value \\
    \hline
    $\log_{10}\epsilon_{1500}^{z=0}$  &  25.63 & 
    $\gamma_{1500}$ & 2.06  \\
    $\alpha_{1500}^{z=0}$ & -0.08 &
    $C_{1500}$ & 1.85  \\
    $\alpha_{1100}^{z=0}$ & -3.71 &
    $C_{1100}$ & 0.50 \\
    $\alpha_{900}^{z=0}$ & -1.5 & 
    ${\rm EW}^{z=0.3}$ & -6.17\,\AA \\
    ${\rm EW}^{z=1}$ & 88.02\,\AA &
    ${\rm EW}^{z=2}$ & 176.7\,\AA  \\
    $\log_{10}f_{\rm LyC}^{z=1}$ & -0.53 &
    $\log_{10}f_{\rm LyC}^{z=2}$ & -0.84  \\
    \hline
    $b_{1500}^{z=0}$ & 0.32 & 
    $\gamma_{\nu}$ & -0.86  \\
    $\log_{10}[\epsilon b]_{1500}^{z=0}$ & 25.62 &
    $\gamma_z$ & 0.79 \\
    \hline
    \end{tabular}
    \smallskip 
    \caption{Parameters to model the emissivity $\epsilon(\nu,z)$ in Eqs.~\eqref{eq:epsilon_1},~\eqref{eq:epsilon_2},~\eqref{eq:epsilon_3} and the bias in Eq.~\eqref{eq:bias}; in the analysis in Sect.~\ref{sec:analysis}, we use the parameter $\log_{10}[\epsilon b]_{1500}^{z=0}$ to collect the degeneracy between $\epsilon_{1500}^{z=0}$ and $b_{1500}^{z=0}$. Values from~Ref.~\cite{Chiang:2018miw}.}
    \label{tab:pars}
\end{table}

\subsection{Observed intensity}\label{sec:intensity}

The volume emissivity we modeled so far describes the UV emission at the source; broadband observations, however, have to deal with a combination of effects to convert this quantity into the observed, integrated, specific intensity $J_{\nu_{\rm obs}}\,[{\rm Jy/sr}]$.

First of all, the presence of HI clumps along the line-of-sight partially absorbs the UV radiation before it is  observed. The effective optical depth $\tau(z)$ of this phenomenon has been originally estimated in~Ref.~\cite{Madau:1995js}, accounting for a Poisson distribution of the absorbers as a function of their column density and redshift. Absorption is efficient for radiation with $\lambda_{\rm rest}$ either smaller than the ionization wavelength 912\,\AA \, or comparable with Ly$\alpha$ and other lines in the Lyman series, which give rise to the Lyman forest~\cite{Gunn:1965,Steidel:1987,Madau:1995js}.
In our analysis, we rely on the improved semi-analytical model in~Ref.~\cite{Inoue:2014zna}, described by a piecewise power-law in overall agreement with~Ref.~\cite{Madau:1995js}. 

The UV volume emissivity in our computation hence is given by $\epsilon(\nu,z)e^{-\tau(\nu)}$, where $\tau(\nu)$ is the optical depth.
By using the radiative transfer function in an expanding Universe, it is possible to convert this quantity into specific intensity~\cite{Gnedin:1996qr}, through
\begin{equation}\label{eq:jnu}
    j(\nu_{\rm obs}) = \frac{c}{4\pi}\int_0^\infty \frac{dz}{H(z)(1+z)}\,\epsilon(\nu,z)\,e^{-\tau(\nu)}.
\end{equation}
where $\nu_{\rm obs} = \nu/(1+z)$, $c$ is the speed of light and $H(z)$ the Hubble factor. In a broadband survey, such as ULTRASAT or GALEX, observations are performed over a wide range of frequencies, weighted by the detector response function. The observed monopole, hence, is  
\begin{equation}\label{eq:intensity}
    {J}_{\nu_{\rm obs}} = \int_{\nu_{\rm obs}^{\rm min}}^{\nu_{\rm obs}^{\rm max}} \frac{d\nu_{\rm obs}}{\nu_{\rm obs}}\,j(\nu_{\rm obs}){R}(\nu_{\rm obs})\,,
\end{equation}
where $R(\nu_{\rm obs})$ is the quantity in Eq.~\eqref{eq:response} once the wavelength is converted into frequency. We estimate $ {J}_{\nu_{\rm obs}} = 237\,{\rm Jy/sr}$ for ULTRASAT, while $ {J}_{\nu_{\rm obs}} = 264\,{\rm Jy/sr},\,79\,{\rm Jy/sr}$ for GALEX NUV and FUV respectively, in agreement with measurements presented in \hyperlink{C19}{\color{magenta} C19}. 

Our modeling of $J_{\nu_{\rm obs}}$ only accounts for extragalactic contributions, while real data include foregrounds. As we will discuss in the next section, the CBR technique generally provides an unbiased estimator of the extragalactic radiation field, which overcomes the foreground problem.

\noindent
At this stage, we can thus safely rely on a signal model that only contains extragalactic UV contributions, and  include the foregrounds only in the noise budget.

\subsection{Spectral tagging via clustering-based redshifts}\label{sec:CBR}

The band-averaged specific intensity in Eq.~\eqref{eq:intensity} contains the integrated information about the redshift evolution of the UV volume emissivity. The CBR technique~\cite{Newman:2008mb,McQuinn:2013ib,Menard:2013aaa} offers a possible way to reconstruct the redshift dependence $dJ_{\nu_{\rm obs}}/dz$, based on the correlation between the broadband survey and the spectroscopic galaxy catalog that is used as reference sample.

On one side, the specific intensities $J_{\nu_{\rm obs}}$ measured in the pixels of the broadband map provide a dataset with unknown redshifts, whose spatial fluctuations are determined by the way UV emitters trace the underlying large-scale structure (LSS). On the other side, galaxies in the spectroscopic sample trace the LSS as well, and they can be divided in consecutive slices of well-known redshift. The angular cross correlation between the two datasets can then be exploited to remap $J_{\nu_{\rm obs}}$ in terms of the redshift-dependent specific intensity
\begin{equation}\label{eq:dJnu}
\begin{aligned}
    \frac{dJ_\nu(z)}{dz}=&\frac{c}{4\pi H(z)(1+z)}\,\,\times\\
    &\times \int_{\nu_{\rm obs}^{\rm min}}^{\nu_{\rm obs}^{\rm max}}\frac{d\nu_{\rm obs}}{\nu_{\rm obs}}R(\nu_{\rm obs})\epsilon(\nu,z)e^{-\tau(\nu_{\rm obs},z)}\,.
\end{aligned}
\end{equation}
The amplitude of the correlation, in fact, will peak for intensity-galaxy pairs who trace the LSS in a common redshift range. As \hyperlink{C19}{\color{magenta} C19} shows, this quantity can be constructed by using the absolute value of the intensity, namely including foregrounds and Galactic emissions; even if this is the case, the cross correlation is capable of isolating the extragalactic contribution. 

The measurement obtained with CBR, however, will be degenerate with the bias of the sources of the unknown-redshift sample, namely the UV emitters; following \hyperlink{C19}{\color{magenta} C19} and \hyperlink{S21}{\color{magenta} S21}, we parameterize it as \begin{equation}\label{eq:bias}
    b(\nu,z) = b_{1500}^{z=0}\left[\frac{\nu}{\nu_{1500}}\right]^{\gamma_{b_\nu}}(1+z)^{\gamma_{b_z}},
\end{equation}
where we normalized the frequency dependence to the value at 1500\,\AA, analogously to the other parameters used to model the emissivity in Sect.~\ref{sec:emissivity}; fiducial values are collected in Table~\ref{tab:pars}. The value of $b_{1500}^{z=0}$ is degenerate with $\epsilon_{1500}^{z=0}$ in determining the amplitude of the signal, as will become evident in the following.

In the data of the UV broadband survey, the information on the clustering enters through an effective intensity-weighted bias $b_J(z)$, which is computed from the source bias in Eq.~\eqref{eq:bias} through
\begin{equation}\label{eq:bias_J}
b_J(z) = \frac{\int_{\nu_{\rm obs}^{\rm min}}^{\nu_{\rm obs}^{\rm max}} {d\nu_{\rm obs}}{\nu_{\rm obs}^{-1}}\,{R}(\nu_{\rm obs})\,b(\nu,z)\epsilon(\nu,z)\,e^{-\tau(\nu)}}{\int_{\nu_{\rm obs}^{\rm min}}^{\nu_{\rm obs}^{\rm max}} {d\nu_{\rm obs}}{\nu_{\rm obs}^{-1}}\,{R}(\nu_{\rm obs})\epsilon(\nu,z)\,e^{-\tau(\nu)}}.
 \end{equation}

To model the CBR observable, we start by defining the fluctuations in the map measurements in terms of the angular position $\phi$ as 
\begin{equation}
    \Delta J_{\nu_{\rm obs}}(\phi) = J_{\nu_{\rm obs}}(\phi) - \langle J_{\nu_{\rm obs}}\rangle,
\end{equation}
where $\langle J_{\nu_{\rm obs}}\rangle$ is the ensemble average. The galaxy survey, instead, is used to trace the 3D overdensity field through
\begin{equation}
    \delta_g(\phi,z) = \frac{n_g(\phi,z)-\langle n_g(z)\rangle}{\langle n_g(z)\rangle} = b_g(z)\delta_m(\phi,z),
\end{equation}
where $n_g(\phi,z)$ is the galaxy number density, $b_g(z)$ the galaxy bias and $\delta_m(\phi,z)$ describes the LSS overdensities in the underlying Dark Matter (DM) field. 

\begin{figure}[ht!]
    \centering
    \includegraphics[width=\columnwidth]{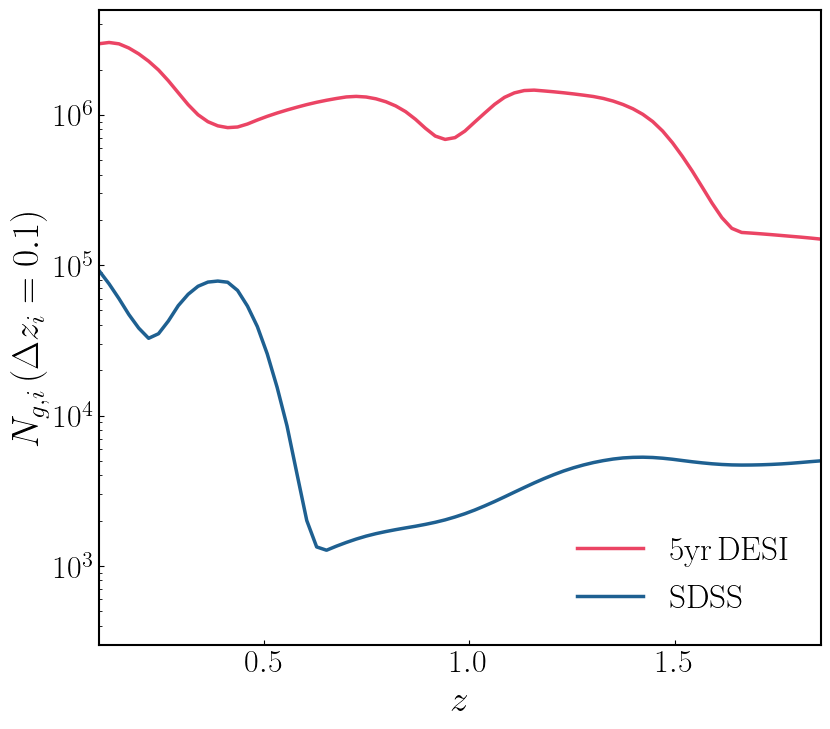}
    \caption{Number of galaxies per redshift bin with size $\Delta z = 0.1$; values for DESI are from~Ref.~\cite{DESI:2023dwi}, while SDSS from Refs.~\cite{SDSS:2004dnq,Reid:2015gra}.  Discontinuities are due to different catalogs.}
    \label{fig:gals}
\end{figure}

The choice of the reference galaxy survey depends on the redshift range and precision one wants to achieve. \hyperlink{C19}{\color{magenta} C19} uses SDSS data from~Refs.~\cite{SDSS:2004dnq,Reid:2015gra}, while we decided to adopt the forecasts for the 5-year results of DESI, described in~Ref.~\cite{DESI:2023dwi}. As Table~\ref{tab:surveys_galaxy} shows, both  surveys provide catalogs mapping different kind of sources; Fig.~\ref{fig:gals} shows the number of galaxies they contain in each redshift bin, namely 
\begin{equation}
    N_{g,i} = \Delta z_i\Omega_{\rm survey}\frac{dN_g}{dzd\Omega}\,,
\end{equation} 
where $dN_g/dzd\Omega$ is the number of galaxies per steradian per redshift bin, $\Omega_{\rm survey}$ is the sky area observed by the survey, and $\Delta z_i$ the width of the bin.

\begin{table}[ht!]
    \centering
    \addtolength{\tabcolsep}{-0.15em}
    \begin{tabular}{|c|ccc|cc|}
    \hline
     & $N_g$ & $\lambda /\Delta \lambda$
     & $\Delta z_i$ & catalog & $z$ \\
    \hline
     \multirow{4}{*}{DESI} & \multirow{4}{*}{$\mathcal{O}(10^7)$} & \multirow{4}{*}{0.001} & \multirow{4}{*}{0.1} & BGS& [0.1, 0.4]   \\
     & && &LRG & [0.4, 1.1]  \\
     & & &&ELG & [1.1, 1.6]  \\
     & & && QSO & [1.6, 2] \\
    \hline
     \multirow{4}{*}{SDSS}  & \multirow{4}{*}{$\mathcal{O}(10^5)$} & \multirow{4}{*}{0.002}& \multirow{4}{*}{0.1}& NYU MAIN & [0.1, 0.2]   \\
     & && &BOSS LOWZ & [0.1, 0.4]  \\
     & && &BOSS CMASS & [0.4, 0.7]  \\
     & && &DR14 QSO & [0.7, 2]  \\
    \hline
    \end{tabular}
    \smallskip 
    \caption{Specs of the galaxy surveys (DESI,~\cite{DESI:2023dwi}, and SDSS,~\cite{SDSS:2004dnq,Reid:2015gra}) considered in this work. The forecasted sources in DESI are divided between bright galaxies (BGS), luminous red galaxies (LRG), emission line galaxies (ELG) and quasars (QSO); SDSS instead contains results from the galaxy NYU value-added catalog, the luminous red galaxy samples in BOSS and the quasars in the DR14 release. The total number $N_g$ is obtained by integrating $dN_g/dzd\Omega$ over the 
    redshift and observed sky area. $\lambda /\Delta \lambda$ defines the instrumental resolution, which sets the best possible spectroscopic resolution that one can reach, namely ${\rm min}(\delta z_i) = \lambda/\Delta\lambda$. Finally, 
    $\Delta z_i$ describes the width of the redshift bins used in our forecast analysis, see Sect.~\ref{sec:noise}.}
    \label{tab:surveys_galaxy}
\end{table}

The angular cross correlation between intensity pixels and galaxies that are separated by an angle $\theta$ on the sky is finally defined as 
\begin{equation}
    \omega_{Jg}(\theta,z) = \langle \Delta J_{\nu_{\rm obs}}(\phi)\delta_g(\phi,z)  \rangle,
\end{equation}
and then marginalized over $\theta$ in order to get 
\begin{equation}\label{eq:baromega}
    \bar{\omega}_{Jg}(z) = \int_{\theta_{\rm min}}^{\theta_{\rm max}}d\theta\,W(\theta)\omega_{Jg}(\theta,z)\,.
\end{equation}
This is the main observable we are interested in our analysis. The choice of the window function $W(\theta)$ is arbitrary.

\noindent
In analogy with \hyperlink{C19}{\color{magenta} C19} and \hyperlink{S21}{\color{magenta} S21}, we choose $W(\theta) = \theta^{-0.8}/\int \theta^{-0.8}d\theta $. The minimum angular distance above which the cross correlation is performed, is set to $\theta_{\rm min}(z) = \arctan[r_{\rm min}/D_A(z)]$, where $D_A(z)$ is the cosmological angular diameter distance and $r_{\rm min} = 0.5\,{\rm Mpc}$ is a physical scale chosen to avoid strong non-linear clustering on small scales; to provide a more conservative result, in Sect.~\ref{sec:analysis} we also consider the case in which $r_{\rm min} = 1\,{\rm Mpc}^{-1}$.
For $z \in [0.1,2]$ we get $\theta_{\rm min} \in [4,0.97]\,{\rm arcmin}$, larger than $L_{\rm pix}$ in Table~\ref{tab:surveys}. 
The choice of $\theta_{\rm max}$, instead, is set to avoid large fluctuations in the calibration of the map. In our case, ULTRASAT has a wide field-of-view (FoV) $\Omega_{\rm Fov} = 204\,{\rm deg}^2$, across which observational properties vary.~Ref.~\cite{Shvartzvald:2023ofi} shows that the effective PSF is almost constant up to $\sim 4\,{\rm deg}$ from the center of the FoV; therefore, we assume that inside this range calibrations are stable and we run our analysis up to $\theta_{\rm max} = 4\,{\rm deg}$. In the case of GALEX, instead, \hyperlink{C19}{\color{magenta} C19} chooses $\theta_{\rm max}(z) = \arctan[ 5{\,\rm Mpc}/D_A(z)] \in [44,10]\,{\rm arcmin} = [0.7,0.2]\,{\rm deg}$. We adopt the same value when dealing with GALEX.

The angular cross correlation $\bar{\omega}_{\tilde{J}{\rm g}}(z)$ defined in Eq.~\eqref{eq:baromega} represents the CBR observable, and can then be used to infer the redshift evolution of the specific intensity. In fact, we can re-express it as 
\begin{equation}\label{eq:wJg_z}
 \bar{\omega}_{\tilde{J}{\rm g}}(z) = b_J(z)b_g(z)\frac{dJ_{\nu_{\rm obs}}}{dz}\int_{\theta_{\rm min}(z)}^{\theta_{\rm max}(z)}d\theta\,W(\theta) \omega_{m}(\theta,z),
\end{equation}
where $b_J(z)$ and $b_g(z)$ are the biases of the two LSS tracers, and $\omega_{m}(\theta,z)$ is the angular two-point function of the underlying DM field. This can be related to the non-linear matter power spectrum~\cite{Maller:2003eg} via
\begin{equation}
    \omega_{m}(\theta,z) = \frac{1}{2\pi}\int_0^\infty dk\,kP_m(k,z)\mathcal{J}_0(k\theta\chi(z))\frac{dz}{d\chi},
\end{equation}
where $\mathcal{J}_0$ is the Bessel function of the first type, $\chi(z)$ the radial comoving distance and $dz/d\chi = H(z)/c$. We compute the $P_m(k,z)$ DM power spectrum using the public library \texttt{CAMB}\footnote{\url{https://github.com/cmbant/CAMB}}~\cite{Challinor:2011bk}, and we adopt the halofit model prescription from~Ref.~\cite{Mead:2020vgs}.

The observation of $\bar{\omega}_{\tilde{J}{\rm g}}(z)$ then leads to the reconstruction of the redshift evolution of the intensity weighted by the bias, $d\tilde{J}_{\nu_{\rm obs}}/dz = b_J(z)dJ_{\nu_{\rm obs}}/dz$. This in turn represents a summary statistics of the UV emission and absorption across space and time. We show $d\tilde{J}_{\nu_{\rm obs}}/dz$ in Fig.~\ref{fig:model_summary}, for ULTRASAT and GALEX (NUV, FUV), while Fig.~\ref{fig:noise} shows the CBR signal-to-noise ratio of $ \bar{\omega}_{\tilde{J}{\rm g}}(z)$ when their maps are cross correlated with the galaxy reference catalogs from SDSS or DESI. These are the quantities that enter our analysis in Sect.~\ref{sec:analysis}.


\subsection{Noise}\label{sec:noise}

Refs.~\cite{Newman:2008mb,Menard:2013aaa} estimate analytically the uncertainty on the CBR angular cross correlation $\bar{\omega}_{\tilde{J}{\rm g}}(z)$ introduced in Eq.~\eqref{eq:wJg_z}. We follow a similar reasoning to they do, accounting for the fact that the broadband UV data are measured as intensity in pixels rather than point sources, as  is done instead in photometric galaxy surveys.\footnote{\hyperlink{S21}{\color{magenta} S21} computes the noise differently from us. In particular, their forecast analysis is built directly on $d\tilde{J}_{\nu_{\rm obs}}/dz$ instead of $\bar{\omega}_{\tilde{J}g}$, thus their noise estimate refers to this quantity.}

In the broadband survey, we model the monopole in each pixel by summing the EBL from Eq.~\eqref{eq:intensity} with an estimate of the foreground. This provides an offset between the expected and observed $J_{\nu_{\rm obs}}$, that has indeed been observed. The authors of~Ref.~\cite{Akshaya:2018} estimated the observed surface brightness in photon units\footnote{In photon units, the EBL monopole is $89\,{\rm ph\,cm^{-2}s^{-1}sr^{-1}}$\AA$^{-1}$ for FUV, $172\,{\rm ph\,cm^{-2}s^{-1}sr^{-1}}$\AA$^{-1}$ for NUV.} to be $\sim 250\,{\rm ph\,cm^{-2}s^{-1}sr^{-1}}$\AA$^{-1}$ in the FUV band, and $\sim 550\,{\rm ph\,cm^{-2}s^{-1}sr^{-1}}$\AA$^{-1}$ in the NUV band. \hyperlink{C19}{\color{magenta} C19} explains the offset as due to the presence of three main foregrounds: near-Earth airglow and zodiacal light (the latter being relevant only in the NUV band, see~Ref.~\cite{Murthy:2013}), Galactic dust, and a component whose origin is unknown. We account for all these sources adding to the EBL monopole a fractional contribution $\mathcal{A}_{\rm fg}J_{\nu_{\rm obs}}$, so that $J_{\nu}^{\rm obs} = J_{\nu_{\rm obs}}(1+\mathcal{A}_{\rm fg})$. We set $\mathcal{A}_{\rm fg}=1.8$ for GALEX FUV and $2.2$ for GALEX NUV and ULTRASAT, since they observe similar bands.
We assume data are Poisson-distributed over pixels (thus we estimate the variance as the signal mean value), while fluctuations are Gaussian with variance $\sigma_N^2$ from Eq.~\eqref{eq:sigma_N}; the total noise 
hence is
\begin{equation}\label{eq:sigma_J}
\sigma_J^2 = [J_{\nu_{\rm obs}}(1+\mathcal{A}_{\rm fg})]^2 +\sigma_N^2.    
\end{equation}

As described in Sect.~\ref{sec:CBR}, the CBR cross correlation is computed in patches of area $\pi\theta_{\rm max}^2$, where $\theta_{\rm max} = 4\,{\rm deg}$ for ULTRASAT and $\theta_{\rm max} \sim [0.7,\,0.2]\,{\rm deg}$ for GALEX. Therefore, to infer the noise we need to account for all the possible pairs of pixels and reference galaxies per redshift bin that can be created inside the patches. We rescale the total number of pixels $N_{\rm pix}^{\rm tot}$ inside $\pi\theta_{\rm max}^2$ as
\begin{equation}\label{eq:num_px}
    N_{\rm pix} = N_{\rm pix}^{\rm tot}{\pi\theta_{\rm max}^2}/{\Omega_{\rm sky}} ={\pi\theta_{\rm max}^2}/{A_{\rm pix}},
\end{equation}
where we used $N_{\rm pix}^{\rm tot} = \Omega_{\rm sky}/A_{\rm pix}$, $\Omega_{\rm sky}$ being the sky area observed by the broadband survey, 
while $A_{\rm pix}=L_{\rm pix}^2$ the area of the pixels computed using Table~\ref{tab:surveys}. Meanwhile, the number of galaxies per $z$ bin in the same area is
\begin{equation}
    N_{g,i}^{\theta_{\rm max}} = \frac{dN_{g}}{dzd\Omega}{\Delta z_i \pi\theta_{\rm max}^2} = N_{g,i}\frac{\pi\theta_{\rm max}^2}{\Omega_{\rm survey}},
\end{equation}
where $N_{g,i}=\Delta z_i \Omega_{\rm survey}dN_{\rm g}/dzd\Omega$ is the number of reference galaxies in the observed area of the survey $\Omega_{\rm survey}$ in the $i$-th redshift bin having size $\Delta z_i$ (see Fig.~\ref{fig:gals}). 

Finally, we estimate the number of pairs to be
\begin{equation}\label{eq:pairs_cbr}
    N_{\rm pairs} = N_{\rm pix}N_{g,i}^{\theta_{\rm max}} = 
    \frac{N_{g,i}}{A_{\rm pix}}\frac{(\pi\theta_{\rm max}^2)^2}{\rm{min}[\Omega_{\rm survey},\Omega_{\rm sky}]}\,.
\end{equation}
In the case of ULTRASAT$\times$DESI and GALEX$\times$DESI, the observed area is set by the galaxy survey, namely $\Omega_{\rm survey} = 14000\,{\rm deg}^2$~\cite{DESI:2023dwi}. Following the procedure in \hyperlink{C19}{\color{magenta} C19}, instead, for GALEX$\times$SDSS we set $\Omega_{\rm survey} = 5500\,{\rm deg^2}$, which is the size of the overlapping footprints. In analogy to \hyperlink{S21}{\color{magenta} S21} and~Ref.~\cite{Chiang:2019}, for SDSS we set $\Delta z_i = 0.1$, hence we sample the $[0,2]$ redshift range with 20 bins. This width is larger than the uncertainty of the spectroscopic redshift bins $\delta z_i$: in fact, as it has been shown in~Refs.~\cite{Menard:2013aaa,Rahman:2014lfa}, CBR provides a measurement of the redshift which is degenerate with the bias evolution in $z$. This degrades the goodness of the CBR-redshift inference, and it can lead to the overlap between contiguous bins when their width is too small. Moreover, using too small redshift bins would increase the shot noise term encapsulated in $N_{g,i}$. In the case of SDSS, the width $\Delta z_i = 0.1$ allows us to overcome this issue; in the case of DESI, we make the conservative choice of using the same redshift bin width. 

\begin{figure*}[ht!]
    \centering
    \includegraphics[width=\columnwidth]{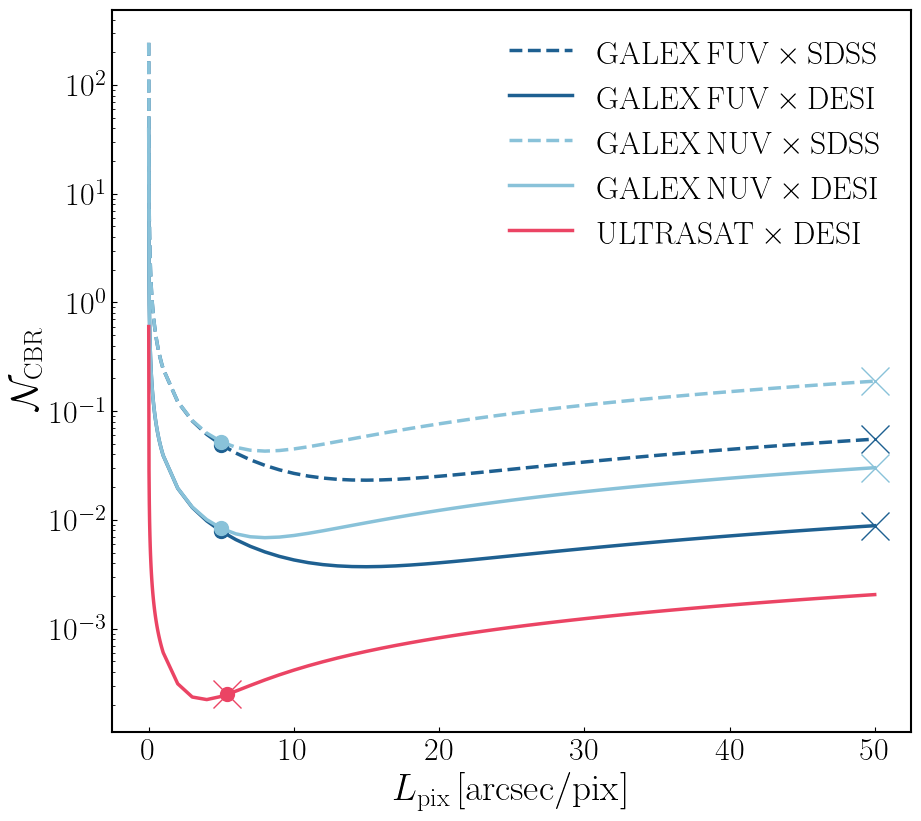}
    \includegraphics[width=\columnwidth]{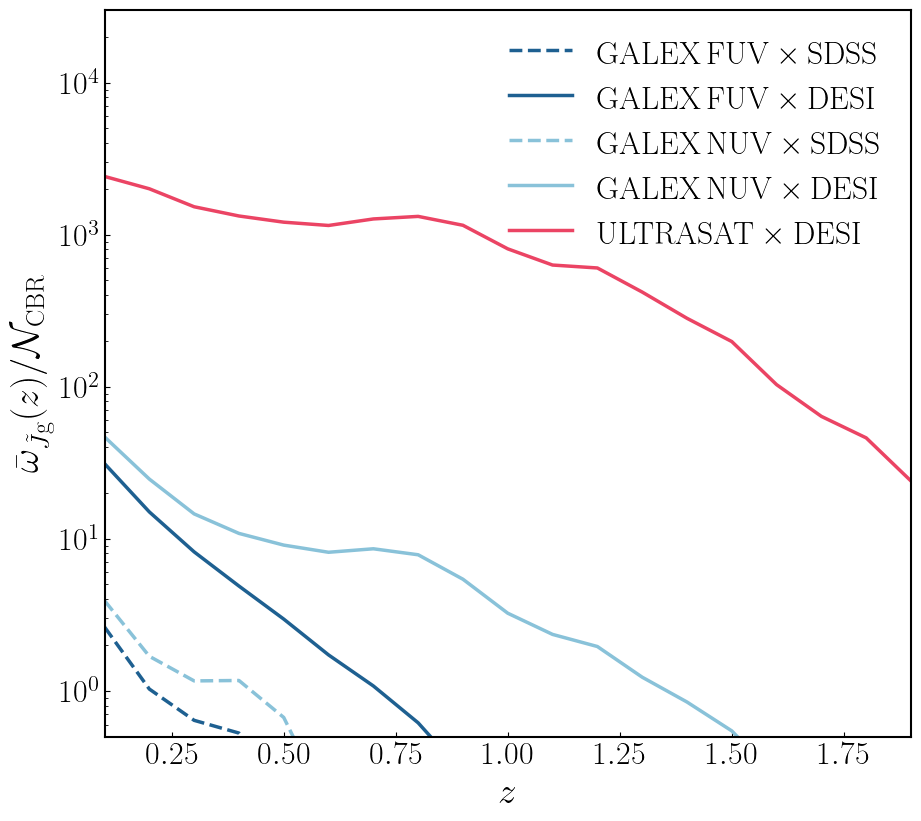}
    \caption{Noise analysis for ULTRASAT$\times$DESI (pink), GALEX NUV$\times$SDSS and $\times$DESI (light blue, dashed and continuous) and  GALEX FUV$\times$SDSS and $\times$DESI (dark blue, dashed and continuous). Left: $\mathcal{N}_{\rm CBR}$ from Eq.~\eqref{eq:noise}. Here, we consider different sizes for the effective pixels; for each detector, the dot indicates the noise for $L_{\rm pix}$ as in Table~\ref{tab:surveys}, while the cross indicates the value adopted in the analysis. Right: SNR obtained from Eqs.~\eqref{eq:wJg_z} and~\eqref{eq:noise}. }
    \label{fig:noise}
\end{figure*}

Moreover, a further contribution to the CBR noise comes from the width $\delta z_i$ of the spectroscopic bins. Assuming that matter clusters on scale $\delta z_c$ and not beyond, a galaxy survey with either bins $\delta z_i >> \delta z_c$ or $\delta z_i<\delta z_c$ would not be precise in reconstructing the $dJ_{\nu_{\rm obs}}/dz$ redshift evolution based on cross correlation information. The best strategy according to~Ref.~\cite{Menard:2013aaa} is to choose reference redshift bins with $\delta z_i \sim \delta z_c\sim 10^{-3}$: in this way, the amplitude of the cross correlation between the intensity and a reference galaxy is high only locally, that is inside the redshift bin from where the intensity comes from. Following~Ref.~\cite{Menard:2013aaa}, we account for this type of noise through a $\delta z_i / \delta z_c$ factor, where we set $\delta z_c = {\rm max}[10^{-3},H(z)(1+z)r_{\rm min}]$ to avoid strong non-linear clustering and account for redshift evolution in the clustering scale. The quantity $r_{\rm min}$ is the same minimum physical distance adopted in Sect.~\ref{sec:CBR} to define the size of the patches in which the cross correlation is computed. On the other hand, 
we define the width of the spectroscopic bins $\delta z_i$ for SDSS and DESI as the largest value between $\delta z_c$ and the instrumental resolution $\lambda/\Delta \lambda$ in Table~\ref{tab:surveys_galaxy}.

Finally, we need to account for the number of sky patches $N_{\theta}$, over which the statistical analysis can be performed. This introduces an extra $(N_\theta)^{-1/2} = \sqrt{\pi\theta_{\rm max}^2/\Omega_{\rm survey}}$ factor. 
Collecting all the elements described up to this point, we estimate the noise in the CBR procedure to be
\begin{equation}\label{eq:noise}
\begin{aligned}
    \mathcal{N}_{\rm CBR} &= \frac{\delta z_i}{\delta z_c}\sqrt{\frac{\sigma_J^2}{N_{\rm pairs}N_{\theta}}} \\
    &= \frac{\delta z_i}{\delta z_c}\sqrt{\frac{A_{\rm pix}{\bigl[[J_{\nu_{\rm obs}}(1+\mathcal{A}_{\rm fg})]^2 + \sigma_N^2\bigr]}}{\pi\theta_{\rm max}^2N_{g,i}}}.
\end{aligned}
\end{equation}
The CBR noise $\mathcal{N}_{\rm CBR}$ clearly depends on the properties of both the broadband and galaxy surveys. To minimize it, on one side we need a spectroscopic survey where the uncertainty $\delta z_i$ is small, but which observes enough galaxies to guarantee a high $N_{g,i}$: the former allows us to improve the quality of the clustering redshift estimation, while the latter reduces the shot noise. On the other side, a sweet spot exists between using small pixels and getting a not-too-large value for $\sigma_N^2$; in fact, using smaller pixels would provide a larger number of pairs, hence a smaller statistical noise; the flux density per pixel $f_\nu^{\rm pix}$ defined in Sect.~\ref{sec:ultrasat}, however, increases when the area of the pixels gets smaller, leading to a larger $\sigma_N^2$. On the other hand, the uncertainty due to the statistical Poissonian fluctuations encapsulated in $J_{\nu_{\rm obs}}(1+\mathcal{A}_{\rm fg})$ determines a ground level for the error budget. Since the noise variance cannot be lower than the monopole expected in the map from the astrophysical contribution, the optimal value is reached when $\sigma_N^2 \sim [J_{\nu_{\rm obs}}(1+\mathcal{A}_{\rm fg})]^2$. By looking at Table~\ref{tab:surveys}, it is evident that ULTRASAT already satisfies this condition with $L_{\rm pix}$, while GALEX requires the use of larger effective pixels. As Fig.~\ref{fig:noise} shows, this translates to two different prescriptions when computing $\mathcal{N}_{\rm CBR}$: while for ULTRASAT the pixel scale $L_{\rm pix}$ is already good enough to minimize the noise, for GALEX it is better to group the pixels, in order to reach an effective scale $L_{\rm pix}^{\rm eff} = 50''/{\rm pix}$, which is indeed the one adopted in \hyperlink{C19}{\color{magenta} C19}. In this case, the noise variance is $\sigma_N^2 = 78\,{\rm Jy/sr}$; finally, the difference between NUV and FUV is due to the different $J_{\nu_{\rm obs}}$. Accurate foreground cleaning and mitigation can reduce the noise with respect to the value we estimated.

Under this choice of parameters, we estimate the noise $\mathcal{N}_{\rm CBR}$ to be used in the analysis in Sect.~\ref{sec:analysis}.
Figure~\ref{fig:noise} compares the signal-to-noise ratio  ${\bar{\omega}_{\tilde{J}{\rm g}}(z)}/\mathcal{N}_{\rm CBR}$ (SNR) estimated for all the surveys; the redshift dependence is evident in the signal, while in the noise it only enters in the choice of $\theta_{\rm max}$ for GALEX and the estimate of $\delta z_i/\delta z_c$.
By looking at GALEX in this plot, it is clear that the cross correlation with DESI will lead to a larger SNR: this is due both to the better spectroscopic redshift determination and to the larger number of galaxies observed. The comparison between GALEX NUV$\times$DESI and ULTRASAT$\times$DESI in this figure reflects instead the smaller noise ULTRASAT has thanks to its smaller $A_{\rm pix}$ and wider $\theta_{\rm max}$. 

\section{Forecasts}\label{sec:analysis}

To constrain the parameters of the UV-EBL emissivity, \hyperlink{C19}{\color{magenta} C19} analyses data from GALEX$\times$SDSS. Since our final goal is to perform a forecast analysis, we rely instead on the Fisher matrix formalism~\cite{Vogeley:1996,Tegmark:1996bz}. 
For each detector-galaxy survey pair, we compute the Fisher matrix by summing over the $i$-redshift bins of the CBR analysis as 
\begin{equation}\label{eq:fisher}
    F_{\alpha\beta} = \sum_{i}\frac{\partial \bar{\omega}_{\tilde{J}g}(z_i)}{\partial\vartheta_\alpha}\frac{\partial \bar{\omega}_{\tilde{J}g}(z_i)}{\partial\vartheta_\beta}\frac{1}{\mathcal{N}_{\rm CBR}^2(z_i)},
\end{equation}
where the noise $\mathcal{N}_{\rm CBR}(z_i)$ is estimated in Eq.~\eqref{eq:noise}. 

\noindent
The parameter set is
\begin{equation}\label{eq:vartheta}
\begin{aligned}
    \vartheta=&\{\log_{10}[b\epsilon]_{1500}^{z=0},\,\gamma_{1500}, \,\gamma_{\nu},\,\gamma_z\\
    & \alpha_{1500}^{z=0},\,C_{1500},\,\alpha_{1100}^{z=0},\,C_{1100},\,\alpha_{900}^{z=0},\\
    &{\rm EW}^{z=z_{\rm EW1}},\,{\rm EW}^{z=z_{\rm EW2}},\\
    &\log_{10}f_{\rm LyC}^{z=z_{\rm C1}},\,\log_{10}f_{\rm LyC}^{z=z_{\rm C2}}\},
\end{aligned}
\end{equation}
which collects the parameters introduced in Table~\ref{tab:pars}. 
Whenever we want to account for priors on certain parameters, we sum the Fisher matrix with the prior matrix, which has $1/\sigma_{\rm prior}^2$ diagonal elements.
The marginalized error on each $\vartheta_\alpha$ can be finally estimated from the inverse of the Fisher matrix (already summed with the prior matrix, if needed) as $\sigma_\alpha = \sqrt{F^{-1}_{\alpha\alpha}}$; these forecasts can then be propagated to estimate the uncertainty on the volume emissivity $\epsilon(\nu,z)$ or on the ionizing photon escape fraction $f_{\rm LyC}$ using
    $\sigma_\aleph = \sqrt{F\mathcal{D}_{\aleph,\vartheta}}$,
where $\aleph$ is the new parameter, $\mathcal{D}_{\aleph,\vartheta} = \left(\partial \aleph/\partial \vartheta_\alpha,...\right)$ the vector of its derivatives with respect to the old ones and  $F$ the Fisher matrix in Eq.~\eqref{eq:fisher}.  

Before proceeding, it is important to remember that, while being extremely useful for order-of-magnitude estimates, the Fisher matrix formalism usually provide optimistic forecasts. In particular, the marginalized errors it obtains tend to underestimate the uncertainty in the presence of significant degeneracies between the parameters; for comparison between forecasts obtained using MCMC and the Fisher approach, see for example~Ref.~\cite{Wolz:2012sr} and references within.

\subsection{Breaking the bias degeneracy}\label{sec:deg_bias}

\begin{table}[ht!]
    \centering
    \addtolength{\tabcolsep}{-0.3em}
    \begin{tabular}{|c|cc|ccc|}
    \hline
         & Conservative & Optimistic & G$\times$S & U$\times$D & (G$+$U)$\times$D \\
    \hline
       $b_{1500}^{z=0}$ & 0.05 & 0.01 & $\checkmark$ & $\checkmark$ & $\checkmark$\\ 
       $\gamma_{b_\nu}$ & 1.30 & 0.30 & $\checkmark$ & $\checkmark$ & $\checkmark$\\
       $\gamma_{b_z}$ & 0.30 & 0.10 & $\checkmark$ & $\checkmark$ & $\checkmark$\\
       $\gamma_{1500}$ & 0.30 & $-$ & $\checkmark$ & $\usym{2717}$ & $\usym{2717}$\\
       $C_{1500}$ & 1.50 & $-$ & $\checkmark$ & $\usym{2717}$ & $\usym{2717}$\\
       $C_{1100}$ & 1.50 & $-$ & $\checkmark$ & $\usym{2717}$ & $\usym{2717}$\\
    \hline
    \end{tabular}
    \smallskip 
    \caption{For all cases, we set priors on the bias parameters. 
    GALEX$\times$SDSS~(G$\times$S) also has informative priors on $\{\gamma_{1500},C_{1500},C_{1100}\}$, as in~Ref.~\cite{Chiang:2018miw}. }
    \label{tab:priors}
\end{table}

As discussed in Sect.~\ref{sec:intensity}, the normalization value $\epsilon_{1500}^{z=0}$ and the local bias $b_{1500}^{z=0}$ are degenerate; for this reason, in our analysis we collect them in a single parameter $\log_{10}[\epsilon b]_{1500}^{z=0}$. Moreover, results in \hyperlink{C19}{\color{magenta} C19} and \hyperlink{S21}{\color{magenta} S21} show further degeneracies between the parameters that determine the slope of the bias, namely $\{\gamma_{b_\nu},\gamma_{b_z}\}$, and the ones that characterize how the emissivity depends on the frequency and redshift. 

To overcome these issues, we introduce Gaussian priors on $\{\gamma_{b_\nu},\gamma_{b_z}\}$; in particular, we rely on \hyperlink{C19}{\color{magenta} C19} results and set $\sigma_{\gamma b_\nu} = 1.30,\,\sigma_{\gamma b_z}=0.30$. These are indeed the largest errorbars \hyperlink{C19}{\color{magenta} C19} obtained from the GALEX$\times$SDSS CBR analysis and can thus be interpreted as a conservative choice for our forecasts. To these priors we add the ones \hyperlink{C19}{\color{magenta} C19} uses on $\{\gamma_{1500},C_{1500},C_{1100}\}$. Imposing a prior $\sigma = 1.3$ on $\gamma_{1500}$, and wide Gaussian priors with $\sigma = 1.5$ on $C_{1500},C_{1100}$ is still needed in the case of GALEX$\times$SDSS, while they can be removed when using DESI.

The motivation behind our prior choice relies on the fact that our main targets are the parameters regulating the emissivity, while the bias can potentially be constrained from different probes, such as the resolved sources catalog of the broadband survey itself.
In \hyperlink{C19}{\color{magenta} C19}, such possibility is explored by estimating the bias of the GALEX resolved sources, and rescaling its value to the EBL regime. To do so, they assume that the bias of the resolved sources is larger than the bias of the EBL diffuse light map, due to the flux-limit that allows us to resolve only of the brightest, and thus more clustered, sources. The analysis is done accounting for the information on the GALEX redshift-dependent luminosity threshold and the luminosity-dependent SDSS galaxy bias in~Ref.~\cite{SDSS:2010acc}, and it returns an estimated errorbar $\sigma_{b_{1500}^{z=0}}=0.05$.

We assume that a similar procedure will be applied to ULTRASAT$\times$DESI, and GALEX$\times$DESI; the estimated errorbars in that cases will reasonably be $\leq 0.05$. Similarly to what \hyperlink{C19}{\color{magenta} C19} discusses, this will make it possible to break the degeneracy between $b_{1500}^{z=0}$ and $\epsilon_{1500}^{z=0}$.
Hence, in the following, we present results for $\epsilon(\nu,z)$; these are obtained by marginalizing over $\{b_{1500}^{z=0},\gamma_{b_\nu},\gamma_{b_z}\}$ with Gaussian priors respectively $\{0.05,0.30,1.30\}$. We stress again that these values are chosen according to GALEX$\times$SDSS results, therefore represent a conservative choice in our forecast. To get a more optimistic result, we also test the case $\{0.01,0.10,0.30\}$. Priors are summarized in Table~\ref{tab:priors}.

\subsection{Validation and parameter forecast}\label{sec:galex_post}

To validate our results, first of all we apply the Fisher  
formalism to GALEX FUV$\times$SDSS and NUV$\times$SDSS and we try to ``post-dict" the results in 
\hyperlink{C19}{\color{magenta} C19}. 
We use the same reduced parameter set that they adopted, namely 
\begin{equation}
\begin{aligned}
    \vartheta_{\rm C19}=&\{\log_{10}[\epsilon b]_{1500}^{z=0},\,\gamma_{1500},\alpha_{1500}^{z=0},\,C_{1500},\,\alpha_{1100}^{z=0},\\
    & C_{1100},\,{\rm EW}^{z=0.3},{\rm EW}^{z=1},\,\gamma_{\nu},\,\gamma_z\}.    
\end{aligned}
\end{equation}
Here, we fixed all the parameters related with the ionizing continuum below the Lyman break in Eq.~\eqref{eq:epsilon_3}, for which \hyperlink{C19}{\color{magenta} C19} showed that GALEX$\times$SDSS has no constraining power, and we adopt the priors in Table~\ref{tab:priors}. 

We run the analysis in Eq.~\eqref{eq:fisher} separately for the NUV and FUV filters, and we assume that they are uncorrelated, so we can sum their matrices to improve the constraints. The fact that the filters are sensitive to different ranges is crucial to capture the shape of the redshift-dependent parameters in Eq.~\eqref{eq:parameters_zdep}, better reconstructing $\epsilon(\nu,z)$. A further confirmation of this can be seen in the good results the authors of \hyperlink{S21}{\color{magenta} S21} obtained combining the three CASTOR filters.

\begin{table}[ht!]
    \centering
    \addtolength{\tabcolsep}{-0.25em}
    \begin{tabular}{|c|cc|c|}
    \hline
         & \multicolumn{2}{c|}{\footnotesize GALEX$\times$SDSS} & {\footnotesize ULTRASAT$\times$DESI}  \\
         & {\footnotesize C19} & \footnotesize {This work} & Full-sky \\
         \hline
        $\log_{10}[\epsilon b]_{1500}^{z=0}$ & $25.62_{-0.01}^{+0.01}$ & 0.02  & 0.03 \\
         $\gamma_{1500}$ & $2.06_{-0.30}^{+0.31}$ & 0.30 &  0.39 \\
         $\alpha_{1500}^{z=0}$ & $-0.08_{-0.84}^{+1.28}$ & 1.38  & 1.18 \\
         $C_{1500}$ & $1.85_{-1.28}^{+1.22}$ & 1.50 & 0.02 \\
         $\alpha_{1100}^{z=0}$ & $-3.71_{-0.98}^{+1.34}$ & 4.50 & 1.09\\
         $C_{1100}$ & $0.50_{-1.44}^{+1.46}$ & 1.50 & 0.87 \\
         ${\rm EW}^{z=0.3}$ & $-6.17_{-11.43}^{+12.63}$ & 89.8 & $-$ \\ 
         ${\rm EW}^{z=1}$ & $88.02_{-48.87}^{+51.44}$ & 553 & 1.19 \\
         ${\rm EW}^{z=2}$ & $176.7$ & $-$ & 10.2\\
    \hline
    \end{tabular}
    \smallskip 
    \caption{$1\sigma$ marginalized errors from our ``post-diction" for GALEX$\times$SDSS, compared with results in~Ref.~\cite{Chiang:2018miw}, and with the prediction for ULTRASAT$\times$DESI. We use the $\vartheta_{\rm C19}$ set and we applied the priors described in Table~\ref{tab:priors}. \vspace*{-.5cm}}
    \label{tab:res_validate}
\end{table}

Our results are collected in Table~\ref{tab:res_validate} and compared with the actual results of \hyperlink{C19}{\color{magenta} C19}: our Fisher forecasts are in the same ballpark. However, our analysis lacks in accurately reproduce constraints on the equivalent width parameters ${\rm EW}^{z=1,2}$. The reason for this resides in the fact that these parameters only affect a small portion of the spectrum described in Eq.~\eqref{eq:epsilon_2}, hence they introduce a discontinuity in the analytical model that is more difficult to capture in the computation of the numerical derivatives required in Eq.~\eqref{eq:fisher}.\footnote{See also the discussion on the {\it practicality} of the Fisher matrix compared with MCMC in~Ref.~\cite{Wolz:2012sr}).} Despite this, the equivalent width parameters are not degenerate with the others, thus they do not alter the result on the other marginalized errors estimated from the Fisher matrix computation.

Since our ``post-diction" is reasonable, we proceed by applying it to the ULTRASAT$\times$DESI scenario. As we discussed in Sect.~\ref{sec:emissivity}, in this case we decided to pivot the evolution of the equivalent width EW at higher redshift, namely $z_{\rm EW1} = 1,\,z_{\rm EW2} = 2$. Table~\ref{tab:res_validate} presents the results we found: the constraining power is increased with respect to GALEX$\times$SDSS on all the parameters, except $\log_{10}[\epsilon b]_{1500}^{z=0}$ for which GALEX is helped by the wider redshift range probed by its two filters, NUV and FUV. ULTRASAT sensitivity at $z \sim 1$, instead, leads to a large improvement in the parameters determining the redshift evolution, as well as in the line equivalent width. In the case of ULTRASAT$\times$DESI there is no need of using priors on $\{\gamma_{1500},C_{1500},C_{1100}\}$.
We checked that both GALEX$\times$SDSS and ULTRASAT$\times$DESI have no constraining power on the parameters $\{\alpha_{900}^{z=0},\log_{10}f_{\rm LyC}^{z=z_{\rm C1}},\log_{10}f_{\rm LyC}^{z=z_{\rm C2}}\}$ 
of the ionizing continuum. Moreover, including them in $\vartheta$ worsens the results on the other parameters.

Having shown that ULTRASAT$\times$DESI will improve the constraints, we now turn our attention to the full parameter set (GALEX$+$ULTRASAT)$\times$DESI, which we consider as our baseline in the following sections. 
In this case, the full set $\vartheta$ in Eq.~\eqref{eq:vartheta} can be tested. As Table~\ref{tab:res_full} shows, all the parameters are indeed well constrained; the only degeneracy not yet solved is the one between $\gamma_{1500}$ and $\gamma_{b_z}$; using a prior on one of them, helps constraining the other. 
Results are obtained by computing separately the Fisher matrices for GALEX NUV$\times$DESI, GALEX FUV$\times$DESI and ULTRASAT$\times$DESI, then summing them. The full matrix includes the EW pivot parameters at $z_{\rm EW} = \{0.3,1,2\}$, with $z_{\rm EW} = 1$ constrained by both the detectors, while the others either by GALEX or ULTRASAT. 
\begin{table}[ht!]
    \centering
    \addtolength{\tabcolsep}{-0.2em}
    \begin{tabular}{|c|c||c|c|}
    \hline
         \multicolumn{4}{|c|}{(GALEX + ULTRASAT)$\times$DESI }\\
         \hline
         $\log_{10}[\epsilon b]_{1500}^{z=0}$ & $25.62\pm 0.005$& $\gamma_{1500}$ & $2.06\pm 0.30$ \\
          $\alpha_{1500}^{z=0}$ & $-0.08\pm 0.25$ & $C_{1500}$ & $1.85 \pm 0.02$\\
           $\alpha_{1100}^{z=0}$ & $-3.71\pm 0.44 $ & $C_{1100}$ &$0.50\pm 1.16 $ \\
           $\alpha_{900}$ & $-1.5\pm 5.10$ & ${\rm EW}^{z=0.3}$ & $-6.17\pm 3.05 $ \\
           ${\rm EW}^{z=1}$ & $88.02\pm 0.51 $ &
           ${\rm EW}^{z=2}$ & $176.7\pm 2.79 $ \\
$\log_{10}f_{\rm LyC}^{z=1}$ & $-0.53 \pm 0.20 $ & $\log{10}f_{\rm LyC}^{z=2}$ & $-0.84 \pm 0.18$ \\
    \hline
    \end{tabular}
    \smallskip 
    \caption{Forecasted $1\sigma$ marginalized errors from our Fisher analysis for (GALEX+ULTRASAT)$\times$DESI. We consider the full set $\vartheta$ with priors as in Table~\ref{tab:priors}. }
    \label{tab:res_full}
\end{table}

\begin{figure*}[ht!]
    \centering
    \includegraphics[width=\columnwidth]{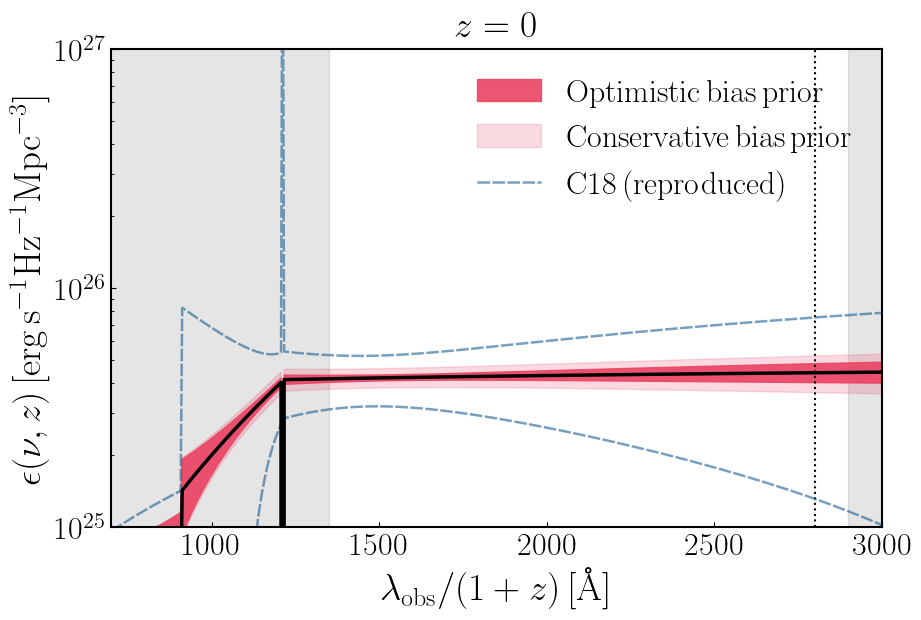}
    \includegraphics[width=\columnwidth]{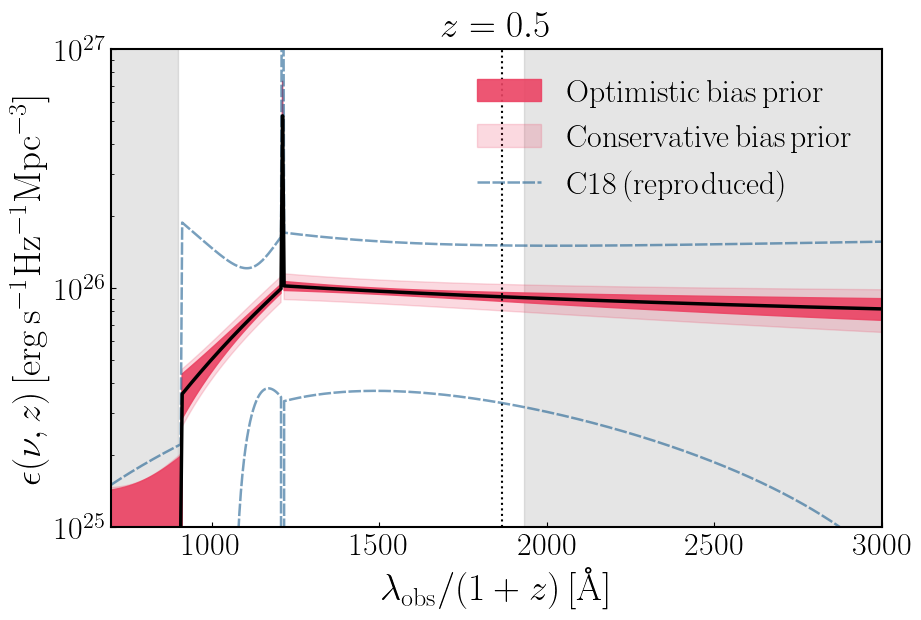}\\
    \includegraphics[width=\columnwidth]{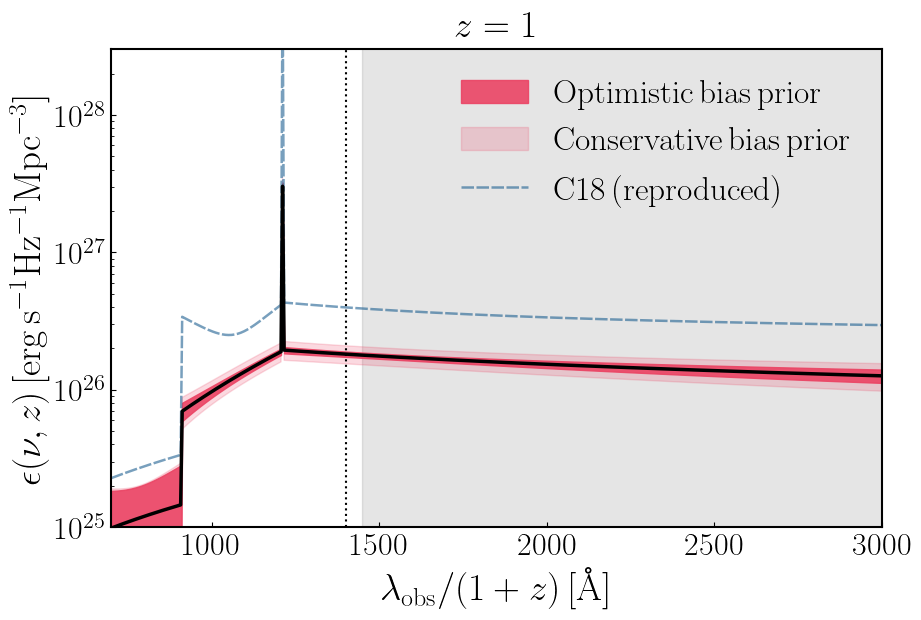}
    \includegraphics[width=\columnwidth]{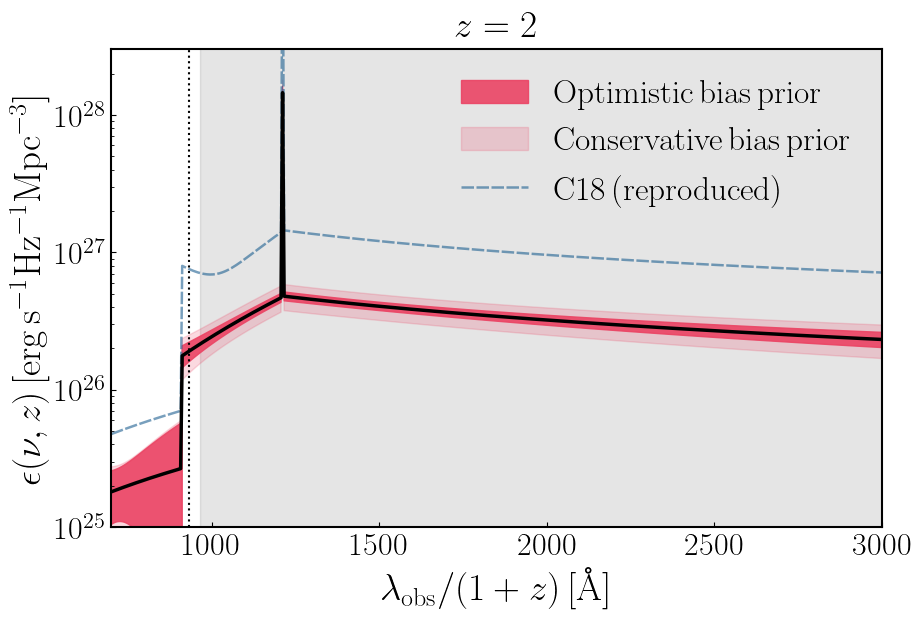}
    \caption{Forecasts at $1\sigma$ on the UV-EBL emissivity $\epsilon(\nu,z)$ at different redshifts for (GALEX$+$ULTRASAT)$\times$DESI, compared with our ``post-diction" at $1\sigma$ for GALEX$\times$SDSS as in~Ref.~\cite{Chiang:2018miw} (compare with Fig.~6 in their work). We tested both the optimistic and conservative bias priors described in Table~\ref{tab:priors}. The gray shaded area delimits the observational window of GALEX$+$ULTRASAT; the gray vertical line shows where GALEX NUV would stop, if ULTRASAT was not included. The parts of the spectrum inside the observed window are directly probed via the CBR, while the others are model-based extrapolations.}
    \label{fig:emissivity_constraints}
\end{figure*}

Our results depend on the choice of $r_{\rm min}=0.5\,{\rm Mpc}$, which we introduced in Sect.~\ref{sec:CBR} to avoid strong non-linear clustering. To test the stability of our conclusions with respect to this parameter, we re-run the Fisher analysis cutting the angular correlation at scales associated with $r_{\rm min} = 1\,{\rm Mpc}$. In this case, the forecasts in Table~\ref{tab:res_full} deteriorate, and the marginalized errors almost double. Despite this, the constraining power remains good even when less optimistic prescriptions are considered. In the next sections, we therefore rely on the constraints obtained in Table~\ref{tab:res_full}; other scenarios can be recovered straightforwardly.

\subsection{Forecasts on the volume emissivity}\label{sec:results_emissivity}

The parameters constrained in the previous section are combined in Eqs.~\eqref{eq:epsilon_1},~\eqref{eq:epsilon_2},~\eqref{eq:epsilon_3} to model the UV-EBL volume emissivity. As discussed in Sect.~\ref{sec:deg_bias}, we can assume the local bias $b_{1500}^{z=0}$ is measured, and so convert the $\log_{10}[\epsilon b]_{1500}^{z=0}$ constraints to $\epsilon_{1500}^{z=0}$ and consequently $\epsilon(\nu,z)$, after marginalizing all the other parameters. We analyse results on $\epsilon(\nu,z)$ using this procedure and applying both the conservative and optimistic bias priors in Table~\ref{tab:priors}.

Figure~\ref{fig:emissivity_constraints} shows our $1\sigma$ forecast constraints on $\epsilon(\nu,z)$ for (GALEX$+$ULTRASAT)$\times$DESI, at $z = \{0,0.5,1,2\}$, using the $\vartheta$ parameter set. Here, we also show how the GALEX$\times$SDSS results from the previous section propagate to $\epsilon(\nu,z)$. Our $1\sigma$ forecasts in this case are obtained using the reduced $\vartheta_{\rm C19}$ and including priors from Table~\ref{tab:priors}; they can be directly compared with Fig.~6 in \hyperlink{C19}{\color{magenta} C19}. 

It is evident that (GALEX$+$ULTRASAT)$\times$DESI will provide very good constraints on the UV-EBL volume emissivity reconstruction. Results could be further improved with respect to our findings, if foreground mitigation was taken into account. For example, setting $\mathcal{A}_{\rm fg}\,=\,0$ in Eq.~\eqref{eq:sigma_J} leads to constraints on the emissivity parameters that are $\sim 0.5$ of the ones in Table~\ref{tab:res_full}.
Figure~\ref{fig:emissivity_constraints} allows us to further comment on an important aspect of our analysis. Not all the redshifted wavelengths of the UV-EBL emissivity fall in the observational windows of GALEX$+$ULTRASAT. The forecast constraints on $\epsilon(\nu,z)$ that are found in the shaded areas are extrapolated knowing the dependencies of the model parameters in Eqs.~\eqref{eq:epsilon_1},~\eqref{eq:epsilon_2},~\eqref{eq:epsilon_3}. The way these have been chosen is agnostic regarding the physics or the type of sources involved: they only require the UV emission to contain a line (Ly$\alpha$), a break (the Lyman break) and a continuum, whose slope varies in the different frequency ranges. 
In the following section, we discuss the implications this will have in our knowledge of the UV-EBL sources.

\subsection{Forecasts on the UV-EBL sources}\label{sec:forecast_sources}

\begin{figure}[ht!]
    \includegraphics[width=\columnwidth]{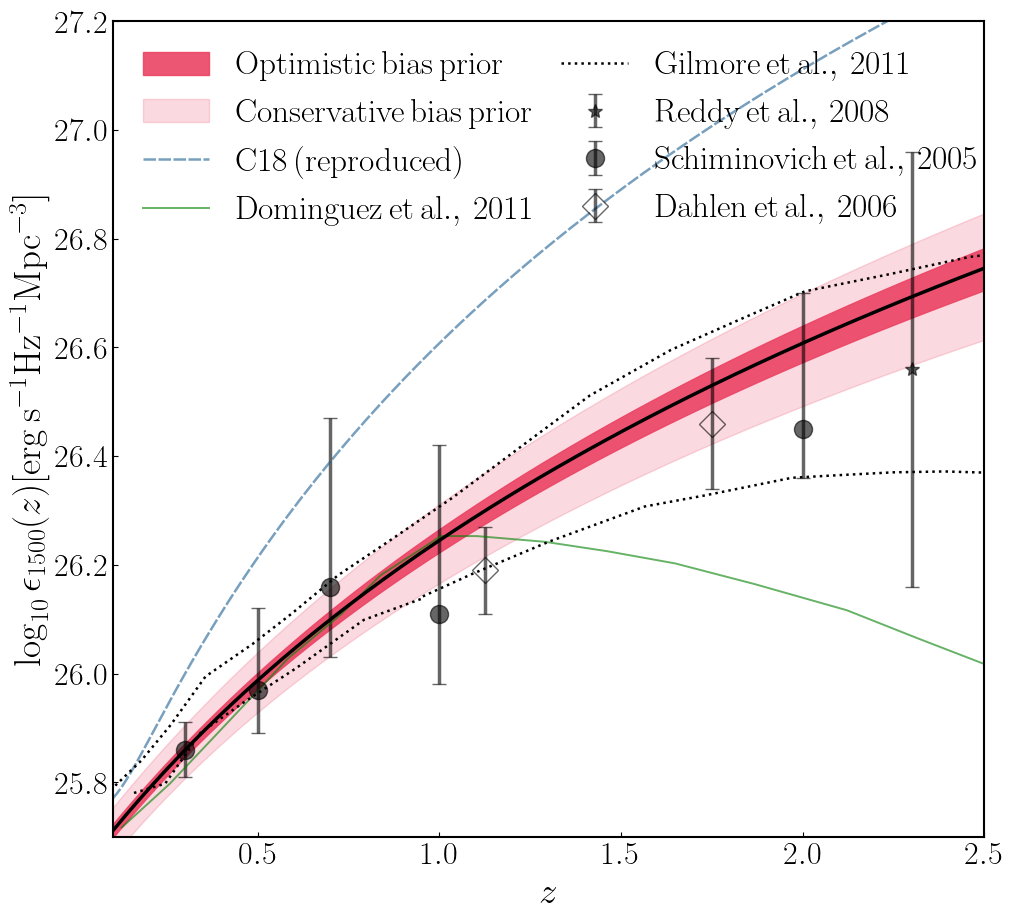}
    \caption{$1\sigma$ forecasts on the non-ionizing $\lambda = 1500\,$\AA\, continuum for (GALEX$+$ULTRASAT)$\times$DESI (shaded area), with conservative and optimistic priors on the bias. To get GALEX$\times$SDSS (C19) results, we propagated our ``post-diction" in Sect.~\ref{sec:galex_post}. We compare with available results in the literature. Ref.~\cite{GALEX-VVDS:2004lih} data are based on the UV luminosity function of the galaxies in GALEX, Ref.~\cite{Dahlen:2006sy} on the GOODS survey from the Hubble Space Telescope, while Ref.~\cite{Reddy:2007bs} use a compilation of data from Hubble. The green line reproduces the fitting of~Ref.~\cite{Dominguez:2010bv} over multiband data, while the dotted lines are the semi-analytical models obtained in~Ref.~\cite{Gilmore:2011ks} with different dust prescriptions. }
    \label{fig:multi_line}
\end{figure}

The shape of the continuum in the non-ionizing region is showed in \hyperlink{C19}{\color{magenta} C19} to be in good agreement with the model in~Ref.~\cite{Haardt:2011xv}, which accounts for galaxy emission and dust extinction.~Ref.~\cite{Haardt:2011xv} stresses that the non-ionizing continuum for instance at 1500\,\AA, provides valuable information on the star formation rate history (modeled e.g.,~in~Ref.~\cite{Madau:1995js}) and the metal-enrichment history (e.g.,~in~Refs.~\cite{Nagamine:2000jb,Kewley:2007}), to be tested against the UV luminosity functions (e.g.,~Refs.~\cite{Wyder:2004ds,Reddy:2008rj,Bouwens:2010gp}). 
Extra-contributions from AGN are negligible in this frequency range, according to~Ref.~\cite{Faucher-Giguere:2019kbp}; on the contrary, the ionizing continuum $\lambda_{\rm rest}< 912\,$\AA\,seems to be AGN-dominated at $z \lesssim 4$. 
The result, however, depends on the AGN luminosity function (e.g.,~Refs.~\cite{Hopkins:2006fq,Kulkarni:2018ebj,Shen:2020obl}) and on the EBL ionizing photon escape fraction $f_{\rm LyC}$, this being largely uncertain.
\hyperlink{C19}{\color{magenta} C19} manages to provide only upper bounds on $f_{\rm LyC}^{z=1},\,f_{\rm LyC}^{z=2}$. In our analysis, we adopted these as our fiducials; actual values of $f_{\rm LyC}$, however, may be smaller than this limit~\cite{Flury:2022asy}. We tested how the observable and forecasts in our analysis change when $f_{\rm LyC}$ is lowered: since this parameter sets the Lyman break depth (compare with Fig.~\ref{fig:model_summary}), it mainly determines the shape of $dJ_{\nu_{\rm obs}}/dz$ at high-$z$, where the signal drops. Therefore, lowering its value only affects the forecasts on $f_{\rm LyC}^{z=z_{\rm C1},z_{\rm C2}}$ and on $\alpha_{900}$: when the escape fraction becomes too small, CBR can no longer constrain the EBL at small wavelengths.

As a final remark, we refer to Fig.~\ref{fig:multi_line} to show that (GALEX$+$ULTRASAT)$\times$DESI will be able to determine the amplitude and redshift evolution of the non-ionizing continuum at $\lambda = 1500\,$\AA\,up to $z\sim 2$, with $1\sigma$ uncertainty $\lesssim 26\%\,(9\%)$ with conservative (optimistic) bias priors, see Table~\ref{tab:priors}. We compare our forecasts with the constraints~Ref.~\cite{GALEX-VVDS:2004lih} obtained using the galaxy UV luminosity function in GALEX, and the ones~Refs.~\cite{Dahlen:2006sy,Reddy:2007bs} got from data compilations from the Hubble Space Telescope. Moreover, we include the fitting~Ref.~\cite{Dominguez:2010bv} realized over multiband Spitzer data, and the semi-analytical models of~Ref.~\cite{Gilmore:2011ks}, which account for different dust models.\footnote{We follow the notation in \hyperlink{C19}{\color{magenta} C19}, where $\epsilon(\nu,z)$ is the comoving specific emissivity; differently from~Ref.~\cite{Haardt:2011xv}, where this symbol indicates the proper volume emissivity. Here, $\epsilon_{1500}(z)$ coincides with the luminosity density, while in~Refs.~\cite{Haardt:2011xv,GALEX-VVDS:2004lih}, is $\rho_{1500}$.}

The improved forecast constrains we obtained on the $\vartheta$ parameters, therefore, can be used to constrain the astrophysical sources of the UV-EBL. With such measurement, the combination of GALEX and ULTRASAT maps will also open the possibility of disentangling any extra-contribution due to emissions beyond galaxies and AGN, for example from decaying DM. We will analyze this enticing possibility in an upcoming, dedicated work.

\section{Conclusion}\label{sec:Conclusion}

Many satellites are planned to be launched and start operating over the coming years. Their characteristics and goals are the most diverse, and it is crucial to understand how to exploit their data as best as we can. New observables and estimators will be needed, in order to probe the Universe at various redshifts and scales, so as to deepen our understanding of cosmic evolution.

Among these space-borne observatories, the Ultraviolet Transient Astronomy Satellite (ULTRASAT,~\cite{Sagiv:2013rma,ULTRASAT:2022,Shvartzvald:2023ofi}) will observe the near-UV (NUV) range between $2300\,$\AA\,and $2900$\,\AA: its main goal will be the study of transients, such as supernovae, variable stars, AGN and electromagnetic counterparts to gravitational-wave sources. To perform its intended analysis, ULTRASAT will first of all build a reference full sky map, which itself will contain an enormous amount of information. In addition, during the lifetime of the mission, its low-cadence survey (LCS) will provide a even deeper map, with sensitivity $\sim 10$ better, covering an area of $\sim 6800\,{\rm deg}^2$.

Besides the emission from resolved galaxies, such map will collect diffuse light of astrophysical and cosmological origin, produced for example by unresolved galaxies, AGN, dust~\cite{Madau:2014bja,Bernstein:2001sq,Haardt:2011xv} or more exotic components such as decaying or annihilating dark matter, or direct-collapse black holes~\cite{Dwek:1998bk,Bond:1985pc,Yue:2012dd,Creque-Sarbinowski:2018ebl,Kalashev:2018bra,Bernal:2020lkd,Carenza:2023qxh}. The overlap of all these processes produce the extragalactic background light (EBL), which in the UV regime is not yet well constrained. 

The ULTRASAT full-sky map will be a good tool to boost our knowledge in this field; to find different ways to exploit its potential, a possible way is to look at the studies that were performed on the full-sky map realized with the All Sky and Medium Imaging Surveys performed by the Galaxy Evolution Explorer (GALEX,~\cite{Martin:2004yr,Morrissey:2007hv}). ULTRASAT, similarly to GALEX, will observe a broad frequency range, hence it will collect the integrated UV emission over a wide redshift range, and map its intensity fluctuations as function of sky position. 

The GALEX diffuse light map has been analyzed by the authors of~Ref.~\cite{Chiang:2018miw} (\hyperlink{C19}{\color{magenta} C19}) via the clustering-based redshift technique (CBR,~Refs.~\cite{Newman:2008mb,McQuinn:2013ib,Menard:2013aaa}): by cross correlating it with the spectroscopic galaxy catalogs from the Sloan Digital Sky Survey (SDSS,~\cite{SDSS:2004dnq,
Reid:2015gra}), they reconstructed the redshift evolution of the comoving volume emissivity of the UV-EBL, providing constraints on the parameters that describe the non-ionizing continuum and the Ly$\alpha$ line. These, in turn, can be used to constrain properties such as the star formation rate or metallicity history. 

A very interesting aspect of CBR is that it is only sensitive to extragalactic contributions; the presence of foregrounds does not alter its signal, while it contributes to the overall noise budget. A similar study was performed in~Ref.~\cite{Scott:2021zue} (\hyperlink{S21}{\color{magenta} S21}) to forecast the constraining power of the Cosmological Advanced Survey Telescope for Optical and UV Research (CASTOR,~\cite{Cote:2019}). CASTOR has been proposed to have three filters, collecting radiation from $\lambda_{\rm obs}\sim 1500$\,\AA\, to $\lambda_{\rm obs}\sim 5500$\,\AA. This will make the analysis sensitive to the EBL in the optical region, as well as to the UV-EBL sourced at higher redshift. The use of three filters will further boost the capability of breaking internal degeneracies between the parameters in the CBR analysis, leading to a better understanding of the EBL cosmic evolution. Further improvements may also come from future experiments that will improve the sensitivity in the FUV band already covered by GALEX; an interesting possibility in this case will be offered by the Ultraviolet Explorer (UVEX,~\cite{Kulkarni:2021tit}).

In this work, we studied how to build forecasts for the CBR technique when this is applied to the cross correlation between a broadband UV survey and a spectroscopic galaxy catalog. We summarized how to model the main observable that is required in this context, namely the angular cross correlation between intensity measurements in pixels and galaxies, and we derived an analytical expression to estimate its noise. We then ran a Fisher forecast with respect to the parameters that model the UV-EBL comoving volume emissivity.

To validate our method, we first of all reproduced the setup of the analysis performed in~\hyperlink{C19}{\color{magenta} C19}, where GALEX$\times$SDSS is considered: the constraints we obtained are in the same ballpark of the actual results. Once we tested the reliability of our analysis, we applied it to forecast the cross correlation between ULTRASAT full-sky maps and galaxies in the spectroscopic bins of the Dark Energy Spectroscopic Instrument (DESI,~\cite{DESI:2013agm,DESI:2016fyo,DESI:2016igz}). We verified that the smaller redshift uncertainty and larger galaxy number density DESI has with respect to SDSS, together with the smaller noise variance and larger field of view in the ULTRASAT map, will imply an improvement in the ULTRASAT$\times$DESI constraints with respect to GALEX$\times$SDSS, when applied to the same parameter set and under the same conditions. 

While in the main text we relied on the ULTRASAT full-sky map, we also analyzed the case of the low-cadence survey. Its $\sim 10$ times better magnitude limit (24.5 vs 23.5) and $\sim 1/2$ smaller field of view (14\,000\,deg$^2$ vs 6\,800\,deg$^2$), lead to forecasts that are $\sim 10\%$ worse than the ones described in Table~\ref{tab:res_full}. We note however that the specifics we assumed for the full-sky map are optimistic, since we do not model in detail the calibration and foreground cleaning procedures. On the other side, the 24.5 magnitude cut for the low-cadence survey represents a conservative choice; reasonably, therefore, the constraints that will be obtained from it will be comparable with our forecasts in Table~\ref{tab:res_full}.

Driven by this result, we applied the forecasted CBR analysis also to the (GALEX$+$ULTRASAT)$\times$DESI full setup. We showed that the large redshift range and the improved sensitivity will allow us to constrain with good accuracy all the parameters in the emissivity model. Once propagated to the emissivity uncertainty, these will lead to the forecasts shown in Fig.~\ref{fig:emissivity_constraints}, where we accounted for different choices of priors on the bias parameters. 

Our results implicitly account for an approximation, namely the uncorrelation between datasets, that allows us to simply sum the Fisher matrices in Eq.~\eqref{eq:fisher}. 
Accounting for correlation between the different datasets is not straightforward: these may arise either because we use the same spectroscopic catalog as reference for the CBR, or because the UV-EBL observed by GALEX FUV, GALEX NUV and ULTRASAT have common sources. To estimate how much this would affect our considerations, we consider an extreme scenario: we assume to separate the reference galaxy catalog in three groups, each of which is cross correlated with only one diffuse light map. Similarly, we assume to mask in each full-sky map $2/3$ of the voxels, to remove double counting of the same sources among different maps. This leads to a factor $\sim 2$ worse constraints with respect to Table~\ref{tab:res_full}.

To further explore this point, we remove from the analysis the Fisher matrix associated with GALEX NUV. Since the FUV filter and ULTRASAT observe non-overlapping bands (see Fig.~\ref{fig:model_summary}), sources that are relevant for the UV-EBL in one of them may not necessarily be relevant for the other. Moreover, using the ULTRASAT map realized in the low-cadence survey map would require only half of the DESI catalog;
we can hence assume to cross correlate GALEX FUV with the other half of its footprint. This would drastically reduce the correlation between the datasets, and it allow us to safely use $F = F_{\rm FUV}+F_{\rm LCS}$. Results are $\lesssim 50\%$ worse than the values collected in Table~\ref{tab:res_full}; even in this case, our technique leads to a good improvement with respect to current constraints on the UV-EBL.

We found that (GALEX$+$ULTRASAT)$\times$DESI will be able to constrain the full UV emissivity frequency dependence in the redshift range $z\lesssim 2$. 
In our work, we assumed that this is composed by different contributions: the non-ionizing continuum, the Ly$\alpha$ line and the ionizing continuum above the Lyman break. The amplitude and slope of each of these elements is determined by the UV-EBL sources; therefore, this measurement will provide interesting insights on the astrophysics. In Fig.~\ref{fig:multi_line}, we showed our forecasts on the non-ionizing EBL emissivity at $\lambda = 1500\,$\AA. Here, current data show an overall agreement with the EBL models that accounts for galaxies, AGN emissions and dust. The improvement 
(GALEX$+$ULTRASAT)$\times$DESI will bring will foster our understanding of this regime, opening the window to the detection of exotic cosmological emissions, if any. 

To conclude, we stress once again how the study of the UV extragalactic background light will be crucial in the upcoming years. ULTRASAT, with its reference map, will offer the possibility of mapping its intensity fluctuations across the full sky, offering a tool to probe the underlying large scale structure in a novel, alternative way. To be able to exploit the enormous amount of scientific information it will contain, it is important to develop specific tools for its analysis; the cross correlation with galaxy surveys, in particular in the context of the clustering-based  redshift analysis, is indeed one of them.

\begin{acknowledgements}
\noindent The authors thank Yi-Kuan Chiang for his comments, which helped improving the quality of the paper. A special thanks goes to José L. Bernal for his insights and suggestions about the analysis, and to the anonymous referee for their comments. We thank the ULTRASAT Collaboration for useful discussion about the project, and Brice M\'enard, Yossi Shvartzvald and Marek Kowalski for feedback on the manuscript. 
SL~thanks the Azrieli Foundation for support and the Padova Cosmology group for hospitality during the tough times between October 7th and November 2023. 
SL is supported by an Azrieli International Postdoctoral Fellowship. EDK was supported by a Faculty Fellowship from the Azrieli Foundation. EDK also acknowledges joint support from the U.S.-Israel Bi-national Science Foundation (BSF,  grant No. 2022743) and the U.S.\ National Science Foundation (NSF, grant No. 2307354), and support from the ISF-NSFC joint research program (grant No. 3156/23). 
\end{acknowledgements}

\bibliographystyle{aa}
\bibliography{biblio}

\end{document}